\documentclass[12pt, draftclsnofoot, onecolumn]{IEEEtran}
\usepackage{amsfonts}
\usepackage{amsmath}
\usepackage{multicol}
\usepackage{stfloats}
\usepackage{color}
\usepackage{}
\usepackage{subfigure}
\makeatletter

\newcommand{\Rmnum}[1]{\expandafter\@slowromancap\romannumeral #1@}
\makeatother
\allowdisplaybreaks[4]

\ifCLASSINFOpdf
  \usepackage[pdftex]{graphicx}
\else
  \usepackage[dvips]{graphicx}
\fi
\begin{document}
%
\title{Jamming-assisted Proactive Eavesdropping over Two Suspicious Communication Links
}

\author{\IEEEauthorblockN{Haiyang Zhang,~\IEEEmembership{Member,~IEEE},~Lingjie Duan,~\IEEEmembership{Senior Member,~IEEE}, and \\Rui Zhang,~\IEEEmembership{Fellow,~IEEE}}
\thanks{


Part of this paper will appear in IEEE International Conference on Communications (ICC), Shanghai, China, 2019 \cite{CP}.

H. Zhang and  L. Duan are with the Engineering Systems and Design Pillar, Singapore University of Technology and Design, Singapore 487372 (e-mail: haiyang\_zhang@sutd.edu.sg;~lingjie\_duan@sutd.edu.sg).

R. Zhang is with the Department of Electrical and Computer
Engineering, National University of Singapore, Singapore 117583 (e-mail: elezhang@nus.edu.sg).

} }

\maketitle

\begin{abstract}

This paper studies a new and challenging wireless surveillance problem where a legitimate monitor attempts to eavesdrop two suspicious communication links simultaneously. To facilitate concurrent eavesdropping, our multi-antenna legitimate monitor employs a proactive  eavesdropping via jamming approach, by selectively  jamming suspicious receivers to lower the transmission rates of the target links. In particular, we are interested in characterizing the achievable eavesdropping rate region for the minimum-mean-squared-error (MMSE) receiver case,  by optimizing the legitimate monitor's  jamming  transmit covariance matrix subject to its power budget.  As the monitor cannot hear more than what suspicious links transmit, the achievable eavesdropping rate region is essentially the intersection of the achievable rate region for the two suspicious links and that for the two eavesdropping links. The former region can be purposely altered by the monitor's jamming transmit covariance matrix, whereas the latter region is fixed when the MMSE receiver is employed. Therefore, we first analytically characterize the achievable rate region for the two suspicious links via optimizing the jamming transmit covariance matrix and then obtain the achievable eavesdropping rate region for the MMSE receiver case.
In addition, as the achievable rate region for suspicious links is in general
non-convex, we also propose a time-sharing based jamming
strategy to enlarge this region and characterize the corresponding
achievable eavesdropping rate region for the MMSE receiver case.
Furthermore, we also extend our study to the MMSE with successive interference cancellation (MMSE-SIC) receiver case and characterize the corresponding achievable eavesdropping rate region by jointly optimizing the time-sharing factor between different decoding orders.
Finally, numerical results are provided to corroborate our analysis and examine the eavesdropping performance.

\end{abstract}

\begin{IEEEkeywords}
Wireless surveillance, suspicious communication, proactive eavesdropping, jamming, beamforming,  interference cancellation.
\end{IEEEkeywords}

\IEEEpeerreviewmaketitle

\section{Introduction}
In recent years, many infrastructure-free wireless communication networks (e.g., device-to-device (D2D) communications, unmanned aerial vehicle (UAV) communications, and mobile ad hoc communications) have emerged as important decentralized supplements to conventional infrastructure-based wireless networks for facilitating direct information exchange among mobile  users \cite{Yong1}-\cite{Add1}. However, these infrastructure-free communication networks may be misused by illegal users such as terrorists, criminals and business spies who deploy a UAV-mounted flying site or an ad hoc link to 
jeopardize public safety, commit crimes, and send back confidential information of trades. In this context, the technique of wireless surveillance was proposed, which allows the legitimate monitor to eavesdrop the suspicious communication links or jam them to disable their communications \cite{Jie1}-\cite{JieA1}.

The existing literature on wireless surveillance can be mainly classified
into two categories. The first one is passive eavesdropping, for which the legitimate monitor silently intercepts the  signals of suspicious links. Note that this approach is efficient only when the channel condition of the eavesdropping link is better than that of the suspicious link. However, in practice, this condition cannot be guaranteed due to sometimes the eavesdropper has to be distant from the suspicious transmitter to avoid getting exposed. To tackle this issue, another approach, namely, proactive eavesdropping, was proposed in \cite{Jie2}-\cite{Jie3}, where the legitimate monitor operates in a full-duplex mode for receiving the suspicious signals and sending jamming signals concurrently. The jamming signals can decrease the data rate of the suspicious link and thus make the eavesdropping feasible even for the case when the eavesdropping link has a worse channel condition than the suspicious link without jamming.  In addition, \cite{Yong2} proposed a spoofing relay approach to enhance the information surveillance capability.
Recently, the proactive eavesdropping approach has  been investigated under various system setups, including relay systems \cite{Relay1}-\cite{Relay4}, multi-antenna systems \cite{MIMO}-\cite{MIMO1}, cognitive radio systems \cite{CR}, UAV-enabled wireless networks \cite{UAV1}-\cite{UAV2}, and parallel fading channels \cite{Yitao}. Specifically, in \cite{Relay1} and \cite{Relay2} the authors studied the wireless surveillance of a two-hop decode-and-forward (DF) relaying communication systems. In \cite{Relay3}, the proactive monitoring via jamming over an amplify-and-forward (AF) relay network was investigated. Different from \cite{Relay1}-\cite{Relay3} in which the relay helps the suspicious transmitter forward messages to the suspicious receiver, \cite{Relay4} considered a proactive eavesdropping scenario with a group of full-duplex AF relays to aid the legitimate monitor for eavesdropping. Moreover, for multi-antenna wireless surveillance systems, efficient jamming schemes were proposed in \cite{MIMO} and \cite{MIMO1} for maximizing the eavesdropping non-outage probability and increasing the worst-case chance  of successful monitoring, respectively. In \cite{CR}, the optimal jamming beamforming vector was designed to maximize the achievable eavesdropping rate in  cognitive radio networks. In \cite{UAV1}, the UAV-assisted wireless surveillance system was investigated, where a full-duplex
monitor located on the ground eavesdrops the ground suspicious communication while sending the collected suspicious information to the UAV. In \cite{UAV2},  an efficient proactive jamming scheme was proposed to maximize the achievable eavesdropping rate in a UAV-aided suspicious communication system. More recently, the authors in \cite{Yitao}  studied the jamming-assisted proactive eavesdropping over parallel independently fading channels.


It should be pointed out that all the above mentioned works consider only the case of a single suspicious communication link. However, in practice, it is likely to have  more than one suspicious communication links in the same district to monitor.  Thus, we are motivated to study a new wireless surveillance scenario where a legitimate monitor eavesdrops two suspicious communication links at the same time. Our legitimate monitor equipped with multiple antennas operates in a full-duplex mode, and tries to intercept and decode the signals from two
suspicious communication links concurrently with the assistance of jamming. Specifically, we focus on characterizing the achievable eavesdropping rate region, by optimizing the legitimate monitor's jamming transmit covariance matrix under the maximum transmit power constraint. The key novelty and main contributions of this paper are summarized as follows.
\begin{itemize}
  \item \emph{Novel proactive eavesdropping over two suspicious communication links via jamming:} To the best of our knowledge, this paper is the first attempt to characterize the achievable eavesdropping rate region over two suspicious links via jamming.  As the legitimate monitor cannot hear more than what suspicious links transmit, we define the achievable eavesdropping rate region as the intersection of the achievable rate region for the suspicious links and that for the eavesdropping links. The former region can be purposely altered by the monitor's jamming signals, whereas the latter region is fixed when the linear minimum mean square error (MMSE) receiver is employed at the legitimate monitor.
  \item \emph{Theoretical characterization of the achievable eavesdropping rate region:}
  We obtain the achievable eavesdropping rate region for the MMSE receiver case by theoretically characterizing the achievable rate region for the suspicious links. As both suspicious links' rates are affected by the legitimate monitor's jamming covariance matrix, we jointly analyze the achievable rate region bounds for both suspicious links. Specifically, we derive the closed-form expressions of the upper and lower boundary points of the achievable rate region for the suspicious links. Moreover, we prove that beamforming is indeed the optimal jamming transmit strategy for the legitimate monitor, and analyze the monotonicity of both the upper and lower boundary curves. Based on these results, we analytically characterize the achievable eavesdropping rate region for the MMSE receiver case. In addition, since the achievable rate region for suspicious links is in general
non-convex, we also propose a time-sharing based jamming
strategy to enlarge this region and characterize the corresponding
achievable eavesdropping rate region.

  \item  \emph{Enlargement of the achievable eavesdropping rate region via  successive interference cancellation:} We also extend our study to the case in which the legitimate monitor applies a non-linear MMSE and successive interference cancellation (MMSE-SIC) receiver to decode the suspicious signals. For this case, we characterize the achievable eavesdropping rate region  by jointly optimizing the time-sharing factor between different decoding orders at the legitimate monitor. Moreover, we derive the optimal time-sharing factor that is sufficient to achieve any point in the achievable eavesdropping rate region for the MMSE-SIC receiver case.

\end{itemize}




It is worth pointing out that the scenario of wireless legitimate surveillance over multiple suspicious links was also considered in \cite{Multi1}, where all suspicious links transmit on orthogonal frequency bands without interfering with each other, and the legitimate monitor has one single receiver antenna and one single transmit antenna. The design criteria in \cite{Multi1} is to maximize weighted sum eavesdropping rate.
Different from \cite{Multi1}, we consider a more complicated multi-antenna surveillance case and focus on characterizing the entire achievable eavesdropping rate region by optimizing the jamming transmit covariance matrix.

\begin{figure}[!ht]
\centering
\includegraphics[scale=1]{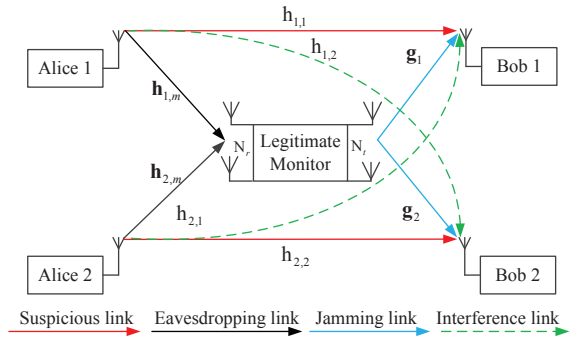}
\caption{Illustration of the legitimate monitor's proactive eavesdropping over two suspicious communication links (Alice 1 to Bob 1 and Alice 2 to Bob 2).}\label{fig:sysmodess}
\end{figure}

The rest of this paper is organized as follows. Section II introduces the system model and defines the achievable eavesdropping rate region for the MMSE receiver case. Section III first studies the achievable rate region for the suspicious links and then characterizes the achievable eavesdropping rate region for the MMSE receiver case. Section IV extends the study to the MMSE-SIC receiver case.  Numerical results are given in Section V. Finally, we conclude the paper in Section VI.

\emph{Notation:}  In this paper, we use boldface lowercase and uppercase letters to denote vectors and matrices, respectively. For a Hermitian matrix ${\bf{A}}$, ${{\bf{A}}^H}$, ${\rm{Tr}}\left( {\bf{A}} \right)$, and ${\rm Rank}\left( {\bf{A}} \right)$ respectively denote its conjugate transpose, trace and rank, while ${{\bf{A}} \succeq 0}$ means that ${\bf{A}}$ is a positive semi-definite matrix.
${{\bf{I}}}$ denotes an identity matrix with an appropriate dimension. $\left\|  {\bf{a}}  \right\|$ computes the Euclidean norm of a complex vector ${\bf{a}}$. For a complex number $z$, $\angle z$ denotes its phase. For any two sets $\mathcal {R}_1$ and $\mathcal {R}_2$, $\mathcal {R}_1\subset \mathcal {R}_2$ (or $\mathcal {R}_1\subseteq \mathcal {R}_2$) denotes that $\mathcal {R}_1$ is a proper subset (or a subset) of $\mathcal {R}_2$, and $\mathcal {R}_1\cap \mathcal {R}_2$ and  $\mathcal {R}_1\cup \mathcal {R}_2$, denote the intersection and union between $\mathcal {R}_1$ and $\mathcal {R}_2$, respectively. All the ${\rm log}\left(\cdot\right)$ functions have base-2 by default.

\section{System Model and problem formulation}

As shown in Fig. \ref{fig:sysmodess}, we consider  proactive eavesdropping over two suspicious communication links, including one legitimate monitor and two suspicious communication pairs. The suspicious transmitters (Alice 1 and Alice 2) and receivers (Bob 1 and Bob 2) are each equipped with one antenna. The legitimate monitor is equipped with $N_r$ receiving antennas for receiving the suspicious messages and $N_t$ transmitting antennas to send jamming signals  for disrupting suspicious receivers. Both the suspicious transmitters send independent suspicious messages to their corresponding receivers over the same frequency band simultaneously.
 The legitimate monitor operates in a full-duplex manner, and we assume that the self-interference  can be perfectly canceled by using advanced analog and digital self-interference cancellation schemes \cite{SIC110}. As will be shown later in Section \Rmnum {5}, the assumption of perfect self-interference cancellation is reasonable in our considered scenario.


Let $h_{i,i}\in {\mathbb{C}^{1 \times 1}}$ denote the channel coefficient of the $i$th suspicious link (from the $i$th suspicious transmitter to the $i$th suspicious receiver, $\forall i\in\{1,2\}$), $h_{i,j}\in {\mathbb{C}^{1 \times 1}}$ denote the channel coefficient of the $i$th interference link (from the $i$th suspicious transmitter to the $j$th suspicious receiver, $\forall i,j\in\{1,2\}$, $j\neq i$),
and  ${\bf h}_{i,m}\in {\mathbb{C}^{N_r\times 1 }}$ denote the channel vector of the $i$th eavesdropping link (from the $i$th suspicious transmitter to the receiver antennas of the legitimate monitor), respectively.
Furthermore, let ${\bf g}_i\in {\mathbb{C}^{N_t \times 1}}$  denote the jamming channel vector from the legitimate monitor to the $i$th suspicious receiver. To characterize the fundamental information-theoretic performance limits of proactive eavesdropping over two suspicious links, we assume that the legitimate monitor has the perfect channel state information (CSI) of all links. In practice, the legitimate monitor can acquire the CSI of ${\bf h}_{i,m}$ and ${\bf g}_{i}$ by overhearing the pilot signals sent by suspicious transmitters and suspicious receivers, respectively. On the other hand, it can obtain the CSI of $h_{i,i}$ and $h_{i,j}$ by eavesdropping the feedback channels of each suspicious transmitter-receiver pair \cite{Yong2}. We also assume that suspicious transmitters are unaware of the existence of legitimate monitor and thus they do not employ any anti-eavesdropping or anti-jamming methods \cite{Jie1}-\cite{Jie2}.

The achievable rate of the $i$th suspicious link is given by
\begin{equation}\label{R1}
R_i^S\left({\bf Q}\right)={\rm log}\left(1+\frac{P_i\left|h_{i,i}\right|^2}{{\rm Tr}\left({\bf Q}{\bf g}_i{\bf g}_i^H\right)+{P_j\left|h_{j,i}\right|^2}+\sigma_i^2}\right), \forall i,j\in\{1,2\}, i\neq j,
\end{equation}
%
where ${\bf Q}\in {\mathbb{C}^{N_t \times N_t}}$  denotes  the jamming transmit covariance matrix at the legitimate monitor, $P_i$  is the transmit power of the $i$th suspicious transmitter, and $\sigma_i^2$ denotes the noise power at the $i$th suspicious receiver. For notation simplicity, we denote ${\tilde \sigma}_i^2={P_j\left|h_{j,i}\right|^2}+\sigma_i^2, \forall i,j\in\{1,2\}, i\neq j$ as the effective noise power at the $i$th suspicious receiver.


We define the achievable rate region for the suspicious links to be
the union of all suspicious rate pairs $\left(R_1^S\left({\bf Q}\right),R_2^S\left({\bf Q}\right)\right)$ over all possible choices of the jamming transmit covariance matrix ${\bf Q} \in \Omega$. Specifically, in the case without (w/o) time-sharing among different jamming transmit covariance matrices, the achievable rate region for the suspicious links is given by
\begin{equation}\label{E9s1}
{\mathcal R}_{\rm w/o-TS}^{\rm S}=\Bigg\{\left(R_1^S\left({\bf Q}\right),R_2^S\left({\bf Q}\right)\right)\Big| {\bf Q} \in \Omega\Bigg\},
\end{equation}
where $\Omega=\left\{{\bf Q} | {\bf Q}\succeq 0, {\rm Tr}\left({\bf Q}\right)\leq P_{\rm max}\right\}$ is the feasible set of ${\bf Q}$, with $P_{\rm max}$ denoting the maximum transmit power at the legitimate monitor. Note that the achievable region ${\mathcal R}_{\rm w/o-TS}^{\rm S}$ also depends on the constant transmission powers, $P_1$ and $P_2$, of the two suspicious transmitters. It is also worth noting that the set ${\mathcal R}_{\rm w/o-TS}^{\rm S}$ is compact because the feasible set $\Omega$ is compact and the mapping from ${\bf Q}$ to $\left(R_1^S\left(\bf Q\right),R_2^S\left(\bf Q\right)\right)$ is continuous.



The MMSE receiver is a linear receiver structure, which optimally trades off capturing the energy of the desired signal of interest and canceling the unwanted interference \cite{Tse}. When the legitimate monitor employs the MMSE receiver to decode different suspicious messages, the achievable rate of the $i$th eavesdropping link observed by the legitimate monitor is expressed as
\begin{equation}\label{E311}
R_i^E={\rm log}\left(1+P_i{\bf h}_{i,m}^H\left({P_j{\bf h}_{j,m}{\bf h}_{j,m}^H}+\sigma_m^2{\bf I}\right)^{-1}{\bf h}_{i,m}\right), \forall i,j\in\{1,2\}, i\neq j,
\end{equation}
where $\sigma_m^2$ denotes the noise power at the legitimate monitor.

The achievable rate region for the eavesdropping links, denoted by ${\mathcal R}_{\rm MMSE}^{\rm E}$,  is defined as the set of all suspicious rate pairs $\left(r_1^E,r_2^E\right)$ that can be decodable at the legitimate monitor simultaneously, i.e.,
\begin{equation}\label{E9s}
{\mathcal R}_{\rm MMSE}^{\rm E}=\Bigg\{\left(r_1^E,r_2^E\right)\Big| 0\leq r_1^E \leq R_1^E; 0\leq r_2^E \leq R_2^E\Bigg\},
\end{equation}
which is a rectangle  specified by the origin and the three vertices $\left(0,R_2^E\right)$, $\left(R_1^E, 0\right)$ and $\left(R_1^E, R_2^E\right)$. The rectangle is fixed when the widely used MMSE receiver is employed at the legitimate monitor.

It is worth pointing out that the legitimate monitor can successfully decode the suspicious messages sent by the $i$th suspicious transmitter if the achievable rate of the $i$th eavesdropping link is no smaller than that of the $i$th suspicious link, $\forall i\in\{1,2\}$. Let $x=\left(r_1^S,r_2^S\right)$ in \eqref{E9s1} denote any suspicious rate-pair in the set ${\mathcal R}_{\rm w/o-TS}^{\rm S}$, i.e., $x\in {\mathcal R}_{\rm w/o-TS}^{\rm S}$. If $x$ falls into the set ${\mathcal R}_{\rm MMSE}^{\rm E}$ in \eqref{E9s}, i,e, $x\in{\mathcal R}_{\rm MMSE}^{\rm E}$, the legitimate monitor can decode both links  simultaneously. Otherwise, if the point  $x$ is outside the set ${\mathcal R}_{\rm MMSE}^{\rm E}$, i.e., $x\notin{\mathcal R}_{\rm MMSE}^{\rm E}$, the legitimate monitor could not guarantee to successfully decode the two suspicious messages concurrently.
Therefore, we define the achievable eavesdropping rate region for the MMSE receiver case as the intersection of the regions ${\mathcal R}_{\rm MMSE}^{\rm E}$ and ${\mathcal R}_{\rm w/o-TS}^{\rm S}$, i.e.,
\begin{equation}\label{E9sss11}
{\mathcal R}_{\rm w/o-TS}^{\rm MMSE}=  {\mathcal R}_{\rm MMSE}^{\rm E} \cap { {\mathcal R}}_{\rm w/o-TS}^{\rm S}
\end{equation}




\begin{figure}[!ht]
\centering
\includegraphics[scale=1]{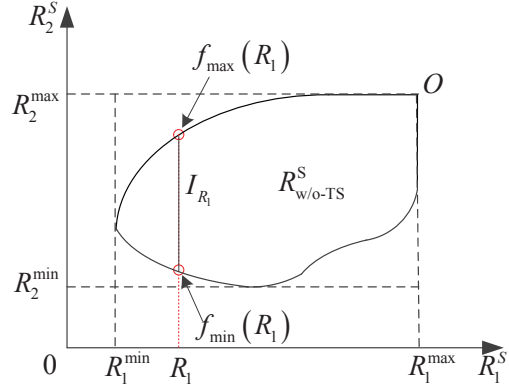}
\caption{An example of the achievable rate region ${\mathcal R}_{\rm w/o-TS}^{\rm S}$ for the two suspicious links. The horizontal and vertical axes represent the achievable rates of Bob 1 and Bob 2, respectively. }\label{201s80725a} 
\end{figure}
In this paper,  we are interested in characterizing the achievable eavesdropping rate region ${\mathcal R}_{\rm w/o-TS}^{\rm MMSE}$ in \eqref{E9sss11}. Note that since the achievable rate region for the eavesdropping links ${\mathcal R}_{\rm MMSE}^{\rm E}$ in \eqref{E9s} is fixed for the MMSE receiver case,  we can derive the achievable eavesdropping rate region ${\mathcal R}_{\rm w/o-TS}^{\rm MMSE}$ by characterizing the achievable rate region for the suspicious links ${\mathcal R}_{\rm w/o-TS}^{\rm S}$ in \eqref{E9s1} only.
Therefore, in the sequel, we focus on characterizing the region ${\mathcal R}_{\rm w/o-TS}^{\rm S}$ driven by the jamming transmit covariance matrix ${\bf Q}$.



\section{Theoretical Characterization of  ${\mathcal R}_{\rm w/o-TS}^{\rm S}$ }
In this section, we focus on characterizing the achievable rate region for the suspicious links  ${\mathcal R}_{\rm w/o-TS}^{\rm S}$. For the ease of exposition, we first give an example of the region ${\mathcal R}_{\rm w/o-TS}^{\rm S}$ in Fig. \ref{201s80725a} and we will introduce more details later.

Prior to characterizing the achievable region ${\mathcal R}_{\rm w/o-TS}^{\rm S}$,  we first need to determine the feasible intervals of
$R_1^S\left({\bf Q}\right)$ and $R_2^S\left({\bf Q}\right)$, respectively. Let $R_1^{\rm min}$  and $R_1^{\rm max}$ denote the minimum and maximum values of
$R_1^S\left({\bf Q}\right)$,  and $R_2^{\rm min}$  and $R_2^{\rm max}$
denote the minimum and maximum values of  $R_2^S\left({\bf Q}\right)$ as shown in Fig. \ref{201s80725a}. Then, we have the following lemma, which confirms the minimum and maximum values of $R_1^S\left({\bf Q}\right)$ and $R_2^S\left({\bf Q}\right)$, respectively.
\newtheorem{lemma}{Lemma}
\begin{lemma}\label{L1107a} $R_1^{\rm min}={\rm log}\left(1+\frac{P_1\left|h_{1,1}\right|^2}{ P_{\rm max}\left\|{\bf g}_1\right\|^2+ {\tilde \sigma}_1^2}\right)$, $R_1^{\rm max}={\rm log}\left(1+\frac{P_1\left|h_{1,1}\right|^2}{ {\tilde \sigma}_1^2}\right)$, $R_2^{\rm min}={\rm log}\big(1+\frac{P_2\left|h_{2,2}\right|^2}{ P_{\rm max}\left\|{\bf g}_2\right\|^2+ {\tilde \sigma}_2^2}\big)$, and $R_2^{\rm max}={\rm log}\left(1+\frac{P_2\left|h_{2,2}\right|^2}{  {\tilde \sigma}_2^2}\right)$.
\end{lemma}
\begin{IEEEproof}
The minimum value of $R_1^S\left({\bf Q}\right)$ is obtained by solving the optimization problem: $R_1^{\rm min}=\min_{{\bf Q} \in \Omega}~~R_1^S\left({\bf Q}\right)$,
which has the same optimal solution as the following problem:
\begin{equation}\label{20181107b}
 \max_{{\bf Q} \in \Omega}~~{\rm Tr}\left({\bf Q}{\bf g}_1{\bf g}_1^H\right).
\end{equation}

According to \cite[Lemma 1]{Jing}, we know that the optimal solution to problem \eqref{20181107b} is ${\bf Q}_{\rm MRT1}={\bf w}_{\rm MRT1}{\bf w}_{\rm MRT1}^H$, where ${\bf w}_{\rm MRT1}=\sqrt{P_{\rm max}}\frac{{\bf g}_1}{\left\|{\bf g}_1\right\|}$ is the maximum-ratio transmission (MRT) beamformer in the direction of ${\bf g}_1$ with the maximum transmit power $P_{\rm max}$. Thus, we have $R_1^{\rm min}=R_1^S\left({\bf Q}_{\rm MRT1}\right)={\rm log}\left(1+\frac{P_1\left|h_{1,1}\right|^2}{ P_{\rm max}\left\|{\bf g}_1\right\|^2+ {\tilde \sigma}_1^2}\right)$.

Similarly, we can obtain that $R_2^{\rm min}={\rm log}\left(1+\frac{P_2\left|h_{2,2}\right|^2}{ P_{\rm max}\left\|{\bf g}_2\right\|^2+ {\tilde \sigma}_2^2}\right)$, which is achievable with the jamming transmit covariance matrix ${\bf Q}_{\rm MRT2}={\bf w}_{\rm MRT2}{\bf w}_{\rm MRT2}^H$, with
${\bf w}_{\rm MRT2}=\sqrt{P_{\rm max}}\frac{{\bf g}_2}{\left\|{\bf g}_2\right\|}$.

Moreover, it is easy to check that both $R_1^S\left({\bf Q}\right)$ and $R_2^S\left({\bf Q}\right)$  are maximized when the legitimate monitor keeps silent and does not send any jamming signals, i,e, ${\bf Q}={\bf 0}$. Thus, we have $R_1^{\rm max}=R_1^S\left({\bf 0}\right)={\rm log}\left(1+\frac{P_1\left|h_{1,1}\right|^2}{ {\tilde \sigma}_1^2}\right)$, and $R_2^{\rm max}=R_2^S\left({\bf 0}\right)={\rm log}\left(1+\frac{P_2\left|h_{2,2}\right|^2}{ {\tilde \sigma}_2^2}\right)$.
\end{IEEEproof}

Note that there exists a rectangle specified by the four vertices $\left(R_1^{\rm min},R_2^{\rm min}\right)$, $\left(R_1^{\rm min},R_2^{\rm max}\right)$, $\left(R_1^{\rm min},R_2^{\rm max}\right)$, and $\left(R_1^{\rm max},R_2^{\rm max}\right)$, which contains the achievable region ${\mathcal R}_{\rm w/o-TS}^{\rm S}$, as shown in Fig. \ref{201s80725a}.

To clearly characterize the achievable region ${\mathcal R}_{\rm w/o-TS}^{\rm S}$,  we next propose an efficient approach to obtain its entire upper and lower boundaries. As illustrated in Fig. \ref{201s80725a}, for each fixed $R_1^S\left({\bf Q}\right)=R_1\in \left[R_1^{\rm min},R_1^{\rm max}\right]$ at Bob 1, we derive the corresponding maximum and minimum achievable rates of Bob 2 (denoted by $f_{\rm max}\left(R_1\right)$ and $f_{\rm min}\left(R_1\right)$, respectively) by solving the following two optimization problems.
\begin{equation}\label{20180903a}
f_{\rm max}\left(R_1\right)=\max_{{\bf Q} \in \Omega}~~R_2^S\left({\bf Q}\right)~~s.t.~~R_1^S\left({\bf Q}\right)={R_1},
\end{equation}
\begin{equation}\label{20180903b}
f_{\rm min}\left(R_1\right)=\min_{{\bf Q} \in \Omega}~~R_2^S\left({\bf Q}\right)~~s.t.~~R_1^S\left({\bf Q}\right)={R_1}.
\end{equation}

It is worth pointing out that the achievable rate-pairs $\left(R_1,f_{\rm max}\left(R_1\right)\right)$ and $\left(R_1,f_{\rm min}\left(R_1\right)\right)$ are, respectively,  the upper and lower boundary points of the region ${\mathcal R}_{\rm w/o-TS}^{\rm S}$ corresponding to any feasible ${R_1}$. Hence, we can find all the upper and lower boundary points of ${\mathcal R}_{\rm w/o-TS}^{\rm S}$ by sweeping ${R_1}$ from $R_1^{\rm min}$ to $R_1^{\rm max}$. In the following two subsections, we focus on solving problems \eqref{20180903a} and \eqref{20180903b} by determining the jamming transmit covariance matrix ${\bf Q}$, respectively.

\subsection{Optimal Solution to Problem \eqref{20180903a} }
In this subsection, we aim to analytically derive the optimal solution to problem \eqref{20180903a} and obtain the closed-form expression of $f_{\rm max}\left(R_1\right)$.

To solve problem \eqref{20180903a}, we first reformulate it as the following equivalent form
\begin{equation}\label{20181029a}
\begin{split}
&\min_{ {\bf Q}\succeq 0}~~{\rm Tr}\left({\bf Q}{\bf g}_2{\bf g}_2^H\right)\\
&~~s.t.~~{\rm Tr}\left({\bf Q}{\bf g}_1{\bf g}_1^H\right)=\phi\left(R_1\right),{\rm Tr}\left({\bf Q}\right) \leq P_{\rm max},
\end{split}
\end{equation}
where $\phi\left(R_1\right)=\frac{P_1\left|h_{1,1}\right|^2}{ 2^{R_1}-1}-{\tilde \sigma}_1^2$ is a nonnegative constant for a fixed $R_1\in \left[R_1^{\rm min},R_1^{\rm max}\right]$.

In fact,  problem \eqref{20181029a} is a semi-definite programming (SDP) problem, which can be solved using the existing numerical optimization toolbox such as CVX \cite{CVX}. However, the numerical approach cannot provide enough insights on the solution structure. To gain more insights of the optimal solution,  we adopt an  analytical approach and first provide the following lemma.
\newtheorem{lemma1107}{Lemma}
\begin{lemma}\label{P20181029a} Let ${\bf{ Q}}^*$ denote the optimal solution to problem \eqref{20181029a}, we then have ${\rm Rank}\left({\bf{ Q}}^*\right) \leq 1$.
\end{lemma}
\begin{IEEEproof}
Notice that problem \eqref{20181029a} is a separable SDP problem with two constraints besides  ${\bf Q}\succeq 0$. According to \cite[Theorem 3.2]{SSDP}, the rank of the optimal solution to problem \eqref{20181029a} should satisfy the inequality constraint: ${\rm Rank}\left({\bf{Q}}^*\right) \leq \sqrt{2}$. Thus, we obtain that ${\rm Rank}\left({\bf{Q}}^*\right) \leq 1$, which completes the proof.
\end{IEEEproof}

According to Lemma \ref{P20181029a}, we know that beamforming is indeed optimal for the legitimate monitor to attain the upper boundary points of the region ${\mathcal R}_{\rm w/o-TS}^{\rm S}$. Thus, we can rewrite ${\bf{Q}}={\bf{w}}{\bf{w}}^H$ with ${\bf{w}}\in {\mathbb{C}^{N_t \times 1}}$. Then, problem \eqref{20181029a} is transformed to
\begin{equation}\label{20180903c}
\begin{split}
\min_{ {\bf w}}~\left|{\bf g}_2^H{\bf w}\right|^2~s.t.~\left|{\bf g}_1^H{\bf w}\right|^2=\phi\left(R_1\right),~\left\|{\bf w}\right\|^2 \leq P_{\rm max}.
\end{split}
\end{equation}

We have the following lemma, which provides the structure of the optimal solution to problem \eqref{20180903c}.

\newtheorem{lemma99}{Lemma}
\begin{lemma}\label{P20180903a}  The optimal jamming beamforming vector ${\bf w}_{\rm opt}$ to problem \eqref{20180903c} is in the form of ${\bf w}_{\rm opt}=\alpha{\bf \hat g}_2+\beta {\bf \hat g}_2^{\perp}$, where $\alpha$ and $\beta$ are two complex weights, ${\bf \hat g}_2=\frac{{\bf g}_2}{\left\|{\bf g}_2\right\|}$, ${\bf \hat g}_2^{\perp}=\frac{{\bf g}_2^{\perp}}{\left\|{\bf g}_2^{\perp}\right\|}$, with ${\bf g}_2^{\perp}=\left({\bf I}-{\bf \hat g}_2{\bf \hat g}_2^H\right){\bf g}_1$ denoting the projection of ${\bf g}_1$ onto the null space of ${\bf \hat g}_2$.
\end{lemma}
\begin{IEEEproof}
Please refer to Appendix A.
\end{IEEEproof}

\begin{figure}[!ht]
\centering
\includegraphics[scale=1]{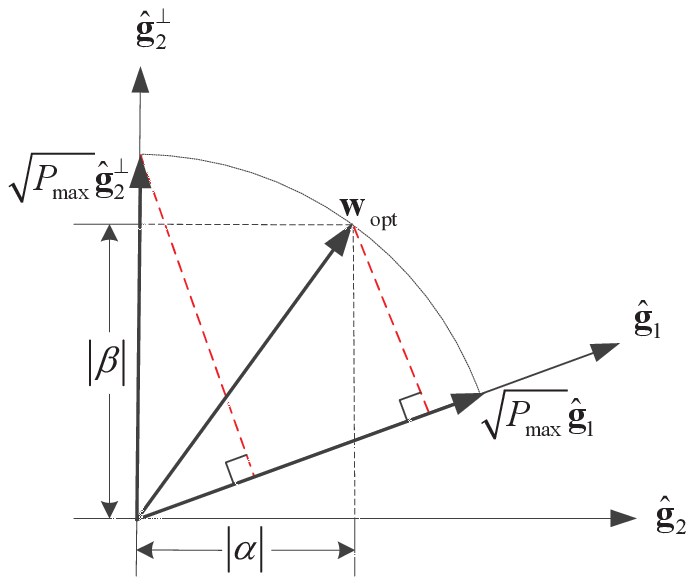}
\caption{The geometric explanation of ${\bf w}_{\rm opt}=\alpha{\bf \hat g}_2+\beta {\bf \hat g}_2^{\perp}$ to problem \eqref{20180903c}. }\label{201s80725sa} 
\end{figure}

Lemma \ref{P20180903a} tells that the optimal jamming beamforming vector to problem  \eqref{20180903c}  should lie in the space spanned by ${\bf \hat g}_2$ and ${\bf \hat g}_2^{\perp}$, as depicted in Fig. \ref{201s80725sa}.
 Then, by substituting ${\bf w}=\alpha{\bf \hat g}_2+\beta {\bf \hat g}_2^{\perp}$ into problem \eqref{20180903c},  it follows that
\begin{equation}\label{20180903g}
\begin{split}
&\min_{\alpha,\beta}~\left|\alpha\right|^2\left\|{\bf g}_2\right\|^2\\
&~s.t.~\left|\alpha{\bf g}_1^H{\bf \hat g}_2\!+\!\beta{\bf g}_1^H{\bf \hat g}_2^{\perp}\right|^2\!=\!\phi\left(R_1\right),\left|\alpha\right|^2\!+\!\left|\beta\right|^2\leq P_{\rm max}.
\end{split}
\end{equation}

It is observed that the objective function of problem \eqref{20180903g} only depends on $\alpha$ and is regardless of $\beta$. Therefore, we need to find the minimum amplitude of $\alpha$ under the condition that the constraints in problem \eqref{20180903g} are not violated. Then, we derive the following theorem, which provides the closed-form solution to problem \eqref{20180903g}.

\newtheorem{theorem}{Theorem}
\begin{theorem}\label{T0906a} The closed-form optimal solution to problem \eqref{20180903g} is given by
\begin{eqnarray}\label{20180903gss}
\left(\alpha^*,\beta^*\right)=
\begin{cases}
\left(0,\sqrt{\frac{\phi\left(R_1\right)}{\left|{\bf g}_1^H{\bf \hat g}_2^{\perp}\right|^2}}e^{-j\angle{\bf g}_1^H{\bf \hat g}_2^{\perp}}\right), &{\rm if~} \phi\left(R_1\right) \leq P_{\rm max}\left|{\bf g}_1^H{\bf \hat g}_2^{\perp}\right|^2\cr \left({\kappa}^*e^{-j\angle{\bf g}_1^H{\bf \hat g}_2},{\iota}^*e^{-j\angle{\bf g}_1^H{\bf \hat g}_2^{\perp}}\right), &{\rm otherwise}\end{cases}
\end{eqnarray}
\end{theorem}
where
${\kappa}^*=\frac{\left|{\bf g}_1^H{\bf \hat g}_2\right|\sqrt{\phi\left(R_1\right)}-\left|{\bf g}_1^H{\bf \hat g}_2^{\perp}\right|\sqrt{\left(\left|{\bf g}_1^H{\bf \hat g}_2\right|^2+\left|{\bf g}_1^H{\bf \hat g}_2^{\perp}\right|^2\right)P_{\rm max}-\phi\left(R_1\right)}}{\left|{\bf g}_1^H{\bf \hat g}_2\right|^2+\left|{\bf g}_1^H{\bf \hat g}_2^{\perp}\right|^2}$ and  ${\iota}^*=\sqrt{P_{\rm max}-{\kappa}^{*^2}}$.
\begin{IEEEproof}
Please refer to Appendix B.
\end{IEEEproof}


The condition $\phi\left(R_1\right) \leq P_{\rm max}\left|{\bf g}_1^H{\bf \hat g}_2^{\perp}\right|^2$ in \eqref{20180903gss} could be re-expressed in terms of $R_1$ as $R_1\geq {\rm log}\left(1+\frac{P_1\left|h_{1,1}\right|^2}{ P_{\rm max}\left|{\bf g}_1^H{\bf \hat g}_2^{\perp}\right|^2+{\tilde \sigma}_1}\right)$. Denote $R_1^{\rm ZF2}= {\rm log}\left(1+\frac{P_1\left|h_{1,1}\right|^2}{ P_{\rm max}\left|{\bf g}_1^H{\bf \hat g}_2^{\perp}\right|^2+{\tilde \sigma}_1}\right)$. Note that $R_1^{\rm ZF2}$ is the value of $R_1$ when jamming signals are sent at the legitimate monitor with the zero-forcing (ZF) beamformer ${\bf w}_{\rm ZF2}=\sqrt{P_{\rm max}}{\hat {\bf  g}}_2^{\perp}$, i.e., transmitting jamming signals in the direction of ${\hat {\bf  g}}_2^{\perp}$ with the maximum transmit power $P_{\rm max}$. Actually, $R_1^{\rm ZF2}$ is the minimum value of $R_1$ that the legitimate monitor can achieve without reducing the achievable rate of Bob 2. Therefore, when $R_1\geq R_1^{\rm ZF2}$ or equivalently $\phi\left(R_1\right) \leq P_{\rm max}\left|{\bf g}_1^H{\bf \hat g}_2^{\perp}\right|^2$, jamming signals are sent only in the direction of ${\hat {\bf  g}}_2^{\perp}$ for minimizing the objective function of \eqref{20180903g}, thus yielding the optimal $\alpha^*=0$ as shown in Theorem \ref{T0906a}. In this case, we also notice that $\left|\beta^*\right|^2=\frac{\phi\left(R_1\right)}{\left|{\bf g}_1^H{\bf \hat g}_2^{\perp}\right|^2}\leq P_{\rm max}$, which means the legitimate monitor may only use part of the transmit power to send jamming signals. Otherwise, if $R_1 < R_1^{\rm ZF2}$ or equivalently $\phi\left(R_1\right) > P_{\rm max}\left|{\bf g}_1^H{\bf \hat g}_2^{\perp}\right|^2$, the legitimate monitor has to send jamming signals in both the directions of ${\hat {\bf  g}}_2 $ and ${\hat {\bf  g}}_2^{\perp}$ with its maximum transmit power budget for achieving $R_1$, resulting in $\alpha^*\neq 0$, $\beta^*\neq 0$, and $\left|\alpha^*\right|^2+\left|\beta^*\right|^2= {\kappa}^{*^2}+{\iota}^{*^2}=P_{\rm max}$. Accordingly, the achievable rate of Bob 2 will be inevitably reduced.

By combining Theorem \ref{T0906a} and Lemma \ref{P20180903a},  we obtain the closed-form optimal solution ${\bf w}_{\rm opt}=\alpha^*{\bf \hat g}_2+\beta^* {\bf \hat g}_2^{\perp}$ to problem \eqref{20180903c}. By substituting ${\bf w}_{\rm opt}$ into the objective function of problem \eqref{20180903a}, we directly obtain the following theorem, which provides the closed-form expression of $f_{\rm max}\left(\gamma_1\right)$.

\newtheorem{theoremss}{Theorem}
\begin{theorem}\label{T1031a} The closed-form expression of $f_{\rm max}\left(R_1\right)$ is given by
\begin{eqnarray}\label{20181003d} f_{\rm max}\left(R_1\right)=
\begin{cases}
{\rm log}\left(1+\frac{P_2\left|h_{2,2}\right|^2}{{\tilde \sigma}_2^2}\right), &{\rm if~}
R_1 \geq R_1^{\rm ZF2}\cr {\rm log}\left(1+\frac{P_2\left|h_{2,2}\right|^2}{ \left({\kappa}^*\right)^2\left\|{\bf g}_2\right\|^2+{\tilde \sigma}_2^2}\right), &{\rm otherwise}\end{cases}
\end{eqnarray}
\end{theorem}

From Theorem \ref{T1031a}, it follows that if $R_1 \geq R_1^{\rm ZF2}$, we have $f_{\rm max}\left(R_1\right)=R_2^{\rm max}={\rm log}\left(1+\frac{P_2\left|h_{2,2}\right|^2}{  {\tilde \sigma}_2^2}\right)$, which indicates the jamming signals sent by the legitimate monitor for reducing the achievable rate of Bob 1 do not decrease that of Bob 2. Otherwise, if  $R_1 < R_1^{\rm ZF2}$, then the achievable rate of Bob 2 will also be reduced by jamming signals.

To examine the characterized region ${\mathcal R}_{\rm w/o-TS}^{\rm S}$ more explicitly, next, we further investigate the monotonicity of  $f_{\rm max}\left(R_1\right)$. Note that $f_{\rm max}\left(R_1\right)$ given in \eqref{20181003d} has two cases, and it is equal to a constant value when $R_1 \geq R_1^{\rm ZF2}$. Thus, in the following, we only study the monotonicity of  $f_{\rm max}\left(R_1\right)$ over the interval $R_1 \in \left[R_1^{\rm min},R_1^{\rm ZF2}\right)$.

\newtheorem{proposition}{Proposition}
\begin{proposition} \label{P20181002a} $f_{\rm max}\left(R_1\right)$ is an increasing function of $R_1 \in \left[R_1^{\rm min},R_1^{\rm ZF2}\right)$.
\end{proposition}
\begin{IEEEproof}
Please refer to Appendix C.
\end{IEEEproof}

Proposition \ref{P20181002a} is also illustrated in Fig. \ref{201s80725a} and  can be intuitively explained as follows. From Fig. \ref{201s80725sa}, it is observed that the amplitudes of the projections of ${\bf w}_{\rm opt}$ onto ${\bf \hat g}_1$ and ${\bf \hat g}_2$, i.e., $\left|{\bf \hat g}_1^H{\bf w}_{\rm opt}\right|$ and $\left|{\bf \hat g}_2^H{\bf w}_{\rm opt}\right|$, are both decreasing when rotating ${\bf w}_{\rm opt}$ anticlockwise from the MRT beamformer ${\bf w}_{\rm MRT1}=\sqrt{P_{\rm max}}{\hat {\bf  g}}_1$ to the zero-forcing (ZF) beamformer ${\bf w}_{\rm ZF2}=\sqrt{P_{\rm max}}{\hat {\bf  g}}_2^{\perp}$. Consequently, in this process, the corresponding rates of Bob 1 and Bob 2 are both monotonically increasing. Thus, Proposition \ref{P20181002a} holds.

\subsection{Optimal Solution to Problem \eqref{20180903b} }
In this subsection, we aim to analytically derive the optimal solution to problem \eqref{20180903b} and provide $f_{\rm min}\left(R_1\right)$ in closed-form.

We first equivalently reformulate problem \eqref{20180903b} as
\begin{equation}\label{20181029b}
\begin{split}
&\max_{ {\bf Q}\succeq 0}~~{\rm Tr}\left({\bf Q}{\bf g}_2{\bf g}_2^H\right)\\
&~~s.t.~~{\rm Tr}\left({\bf Q}{\bf g}_1{\bf g}_1^H\right)=\phi\left(R_1\right),{\rm Tr}\left({\bf Q}\right) \leq P_{\rm max}.
\end{split}
\end{equation}

\begin{figure}[!ht]
\centering
\includegraphics[scale=1]{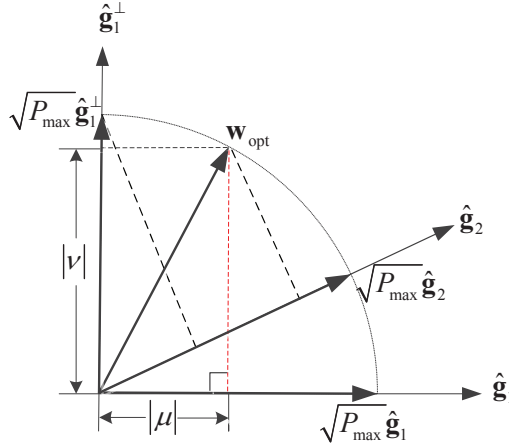}
\caption{The geometric explanation of ${\bf w}_{\rm opt}=\mu{\bf \hat g}_1+\nu {\bf \hat g}_1^{\perp}$ to problem \eqref{20180907a}. }\label{201s80725sadd} 
\end{figure}
Following the same steps as in the proof of Lemma \ref{P20181029a}, we can prove that the optimal solution to problem  \eqref{20181029b} also satisfies ${\rm Rank}\left({\bf{ Q}}^*\right) \leq 1$. Hence, by rewriting ${\bf{Q}}={\bf{w}}{\bf{w}}^H$,
problem \eqref{20181029b} is equivalently reformulated as
\begin{equation}\label{20180907a}
\begin{split}
\max_{ {\bf w}}~\left|{\bf g}_2^H{\bf w}\right|^2~s.t.~\left|{\bf g}_1^H{\bf w}\right|^2=\phi\left(R_1\right),~\left\|{\bf w}\right\|^2 \leq P_{\rm max}.
\end{split}
\end{equation}

Similar to Lemma \ref{P20180903a}, we have the following lemma, which provides the structure of the optimal solution to problem \eqref{20180907a}.
\newtheorem{lemma20181030b}{Lemma}
\begin{lemma}\label{P20180725}  The optimal jamming beamforming vector ${\bf w}_{\rm opt}$ to problem \eqref{20180907a} is in the form of ${\bf w}_{\rm opt}=\mu{\bf \hat g}_1+\nu {\bf \hat g}_1^{\perp}$, where ${\bf \hat g}_1^{\perp}=\frac{{\bf g}_1^{\perp}}{\left\|{\bf  g}_1^{\perp}\right\|}$, ${\bf  g}_1^{\perp}=\left({\bf I}\!-\!{\bf \hat g}_1{\bf \hat g}_1^H\right){\bf g}_2$, $\mu$ and $\nu$ are two complex weights.
\end{lemma}
\begin{IEEEproof}
The proof is similar to that of Lemma \ref{P20180903a}, and thus is omitted for brevity.
\end{IEEEproof}

Lemma \ref{P20180725} indicates that the optimal jamming beamforming vector to problem  \eqref{20180907a}  should lie in the space spanned by ${\bf \hat g}_1$ and ${\bf \hat g}_1^{\perp}$, as depicted in Fig. \ref{201s80725sadd}.
Note that to make problem \eqref{20180907a} easier to solve, we alternatively adopt ${\bf \hat g}_1$ and ${\bf \hat g}_1^{\perp}$ as a set of orthogonal basis vectors to represent ${\bf w}$, rather than using ${\bf \hat g}_2$ and ${\bf \hat g}_2^{\perp}$ as in Lemma \ref{P20180903a}. Then, based on Lemma \ref{P20180725}, we equivalently rewrite problem \eqref{20180907a} as
\begin{equation}\label{E17ssssss}
\begin{split}
&\max_{ \mu, \nu}~~\left|\mu{\bf g}_2^H{\bf \hat g}_1+\nu{\bf g}_2^H{\bf \hat g}_1^{\perp}\right|^2\\
&~~s.t.~~\left|\mu\right|^2\left\|{\bf g}_1\right\|^2=\phi\left(R_1\right),~\left|\mu\right|^2+\left|\nu\right|^2 \leq P_{\rm max}.
\end{split}
\end{equation}

 We then have the following theorem, which provides the closed-form optimal solution to problem \eqref{E17ssssss} and the closed-form expression of $f_{\rm min}\left(R_1\right)$.
\newtheorem{theorem20180725}{Theorem}
\begin{theorem}\label{T20180907ass}  The closed-form optimal solution to problem \eqref{E17ssssss} is
\begin{equation*}\label{E17sssssss}
\mu^*=\sqrt{\frac{\phi\left(R_1\right)}{\left\|{\bf g}_1\right\|^2}}e^{-j\angle{\bf g}_2^H{\bf \hat g}_1},~\nu^*=\sqrt{P_{\rm max}-\frac{\phi\left(R_1\right)}{\left\|{\bf g}_1\right\|^2}}e^{-j\angle{\bf g}_2^H{\bf \hat g}_1^{\perp}},
\end{equation*}
and the resulting optimal value of problem \eqref{20180903b} is given by
\begin{equation}\label{20181001c}
f_{\rm min}\left(R_1\right) ={\rm log}\left(1+\frac{P_2\left|h_{2,2}\right|^2}{\left(\left|\mu^*\right|\left|{\bf g}_2^H{\bf \hat g}_1\right|+\left|\nu^*\right|\left|{\bf g}_2^H{\bf \hat g}_1^{\perp}\right|\right)^2+{\tilde \sigma}_2^2}\right).
\end{equation}
\end{theorem}
\begin{IEEEproof}
Please refer to Appendix D.
\end{IEEEproof}

From Theorem \ref{T20180907ass}, we observe that $\left|\mu^*\right|^2+\left|\nu^*\right|^2=P_{\rm max}$, which means the legitimate monitor needs to use its full transmit power to send jamming signals for minimizing the achievable rate of Bob $2$ given the achievable rate requirement of Bob $1$.

Similar to Proposition \ref{P20181002a}, we have the following proposition,
whose proof is omitted for brevity. This proposition reveals the monotonicity of $f_{\rm min}\left(R_1\right)$ in  \eqref{20181001c}.
\newtheorem{proposition20180707c}{Proposition}
\begin{proposition} \label{20180707c} $f_{\rm min}\left(R_1\right)$ is decreasing in $R_1\in \left[R_1^{\rm min},R_1^{\rm MRT2}\right]$, and is increasing  in $R_1\in \left(R_1^{\rm MRT2},R_1^{\rm max}\right]$, where $R_1^{\rm MRT2}={\rm log}\left(1+\frac{P_2\left|h_{2,2}\right|^2}{ P_{\rm max}\left|{\bf g}_1^H{\bf \hat g}_2\right|^2+ {\tilde \sigma}_2^2}\right)$.
\end{proposition}

Proposition \ref{20180707c} is also illustrated in Fig. \ref{201s80725a} and can be intuitively explained as follows. It is observed from Fig. \ref{201s80725sadd} that the amplitude of the projection of ${\bf w}_{\rm opt}$ onto ${\bf \hat g}_1$, i.e., $\left|{\bf \hat g}_1^H{\bf w}_{\rm opt}\right|$, is decreasing when rotating ${\bf w}_{\rm opt}$ anticlockwise from the MRT beamformer ${\bf w}_{\rm MRT1}$ to the ZF beamformer ${\bf w}_{\rm ZF1}=\sqrt{P_{\rm max}}{\hat {\bf  g}}_1^{\perp}$. In this process, the achievable rate of Bob 1 is thus increasing. However,
with the rotation of ${\bf w}_{\rm opt}$ from ${\bf w}_{\rm MRT1}$ to ${\bf w}_{\rm ZF1}$, the amplitude of the projection of ${\bf w}_{\rm opt}$ onto ${\bf \hat g}_2$, i.e., $\left|{\bf \hat g}_2^H{\bf w}_{\rm opt}\right|$,  is first increasing and then decreasing. Therefore, the resulting achievable rate of Bob 2 is first decreasing and then increasing. Hence, Proposition \ref{20180707c} is established.

\subsection{Characterizing ${\mathcal R}_{\rm w/o-TS}^{\rm S}$}
In the previous two subsections, we have successfully derived the closed-form expressions of $f_{\rm max}\left(R_1\right)$ and $f_{\rm min}\left(R_1\right)$. Thus, the achievable upper and lower  boundary points $\left(R_1,f_{\rm max}\left(R_1\right)\right)$ and $\left(R_1,f_{\rm min}\left(R_1\right)\right)$ of the region ${\mathcal R}_{\rm w/o-TS}^{\rm S}$  (corresponding to any feasible $R_1$) are obtained for Fig. \ref{201s80725a}. Next, we further study whether any point within the two upper and lower boundary points is achievable. Let ${\mathcal {I}}_{R_1}$ denote the line segment connecting these two boundary points. We then have the following proposition, which verifies that any point on
${\mathcal {I}}_{R_1}$ is achievable and provides the required jamming beamforming vector.
\newtheorem{proposition568}{Proposition}
\begin{proposition} \label{20170720aaaa}
Let $\left(R_1,R_2\right)$ be an arbitrary point on the line segment
${\mathcal {I}}_{R_1}$, i.e., $f_{\rm min}\left(R_1\right)\leq R_2 \leq f_{\rm max}\left(R_1\right)$. Then, the point $\left(R_1,R_2\right)$ is achievable with the jamming beamforming vector ${\bf w}_{\rm opt}=\varepsilon_1^*{\bf \hat g}_1+\varepsilon_2^*{\bf \hat g}_1^{\perp}$, with $\varepsilon_1^*=\sqrt{\frac{\phi\left(R_1\right)}{\left\|{\bf g}_1\right\|^2}}e^{-j\angle{\bf g}_2^H{\bf \hat g}_1}$ and $\varepsilon_2^*=\frac{\sqrt{\phi\left(R_2\right)}-\left|\varepsilon_1^*\right|\left|{\bf g}_2^H{\bf \hat g}_1\right|}{\left|{\bf g}_2^H{\bf \hat g}_1^{\perp}\right|}e^{-j\angle{\bf g}_2^H{\bf \hat g}_1^{\perp}}$, where $\phi\left(R_2\right)=\frac{P_2\left|h_{2,2}\right|^2}{2^{R_2}-1}-{\tilde \sigma}_2^2$.
\end{proposition}
\begin{IEEEproof}
The minimum jamming transmit power required to attain the point $\left(R_1,R_2\right)$ is the optimal value of the following optimization problem
\begin{equation}\label{20170720a}
 \begin{split}
 \min_{ {\bf Q}\succeq 0}~{\rm Tr}\left({\bf Q}\right)~s.t.~R_1^S\left({\bf Q}\right)=R_1,~R_2^S\left({\bf Q}\right)= R_2.
 \end{split}
 \end{equation}

Following the same steps as in the proofs of Lemmas \ref{P20181029a} and \ref{P20180903a}, it is easy to verify that beamforming is the optimal jamming transmit strategy (i.e., ${\bf{Q}}={\bf{w}}{\bf{w}}^H$) and the optimal beamforming solution ${\bf w}_{\rm opt}$ is in the form of ${\bf w}_{\rm opt}=\varepsilon_1{\bf \hat g}_2+\varepsilon_1{\bf \hat g}_2^{\perp}$.  Accordingly, problem \eqref{20170720a} is equivalently reformulated as
 \begin{equation}\label{20181101as}
\begin{split}
&\min_{\varepsilon_1, \varepsilon_2}~\left|\varepsilon_1\right|^2+\left|\varepsilon_2\right|^2\\
&s.t.~\left|\varepsilon_1\right|^2\left\|{\bf g}_1\right\|^2\!=\!\phi\left(R_1\right),
\left|\varepsilon_1{\bf g}_2^H{\bf \hat g}_1\!+\!\varepsilon_2{\bf g}_2^H{\bf \hat g}_1^{\perp}\right|^2\!=\!\phi\left(R_2\right),
\end{split}
\end{equation}
 where $\phi\left(R_2\right)=\frac{P_2\left|h_{2,2}\right|^2}{2^{R_2}-1}-{\tilde \sigma}_2^2$.

 Then, similar to the proof of Theorem \ref{T0906a}, the optimal solution to problem \eqref{20181101as} is given by
  \begin{equation*}\label{20181101ass}
  \varepsilon_1^*\!=\!\sqrt{\frac{\phi\left(R_1\right)}{\left\|{\bf g}_1\right\|^2}}e^{\!-\!j\angle{\bf g}_2^H{\bf \hat g}_1}, \varepsilon_2^*\!=\!\frac{\sqrt{\phi\left(R_2\right)}\!-\!\left|\varepsilon_1^*\right|\left|{\bf g}_2^H{\bf \hat g}_1\right|}{\left|{\bf g}_2^H{\bf \hat g}_1^{\perp}\right|}e^{\!-\!j\angle{\bf g}_2^H{\bf \hat g}_1^{\perp}}.
 \end{equation*}

Moreover, considering $f_{\rm min}\left(R_1\right)\leq R_2$ and according to \eqref{20181001c}, we have $\phi\left(R_2\right)\leq \big(\left|\mu^*\right|\left|{\bf g}_2^H{\bf \hat g}_1\right|+ \left|\nu^*\right|\left|{\bf g}_2^H{\bf \hat g}_1^{\perp}\right|\big)^2$. Therefore, we can verify that $\left|\varepsilon_1^*\right|^2+\left|\varepsilon_2^*\right|^2 \leq \left|\mu^*\right|^2+\left|\nu^*\right|^2=P_{\rm max}$. Thus, the point $\left(R_1,R_2\right)$ is achievable within a given jamming power budget $P_{\rm max}$.
 \end{IEEEproof}


\emph{Remark 1:} According to Proposition \ref{20170720aaaa}, we know that each point on the line segment ${\mathcal {I}}_{R_1}$ is achievable. Hence, the entire achievable region ${\mathcal R}_{\rm w/o-TS}^{\rm S}$ can be obtained by sweeping ${R_1}$ from $R_1^{\rm min}$ to $R_1^{\rm max}$.

\subsection{Characterizing ${\mathcal R}_{\rm w/-TS}^{\rm S}$ and ${\mathcal R}_{\rm w/-TS}^{\rm MMSE}$}

In the previous subsections, we have characterized the achievable rate region ${\mathcal R}_{\rm w/o-TS}^{\rm S}$ and analyzed the monotonicity of its boundary curves. As the region ${\mathcal R}_{\rm w/o-TS}^{\rm S}$ is in general non-convex, in this subsection, we propose a
time-sharing-based jamming strategy to further enlarge this region. Specifically, the jamming transmission is divided into two orthogonal time slots and the legitimate monitor is allowed to adopt different jamming beamforming vectors within different time slot.

Let ${\mathcal R}_{\rm w/-TS}^{\rm S}$ denote the achievable rate region for the suspicious links with time-sharing among different jamming beamforming vectors, then we have
\begin{equation}\label{20181126a}
{\mathcal R}_{\rm w/-TS}^{\rm S}
\!=\!\bigcup_{\scriptstyle ~~~~~~0\leq \tau \leq 1 \hfill\atop {\scriptstyle \left(R_1,R_2\right)\in {\mathcal R}_{\rm w/o-TS}^{\rm S} \hfill\atop \scriptstyle \left(R_1^{\prime},R_2^{\prime}\right)\in {\mathcal R}_{\rm w/o-TS}^{\rm S}}}\left(\tau R_1\!+\!\left(1\!-\!\tau\right)R_1^{\prime},\tau R_2\!+\!\left(1\!-\!\tau\right)R_2^{\prime}\right),
\end{equation}
where $\left(R_1,R_2\right)$ and $\left(R_1^{\prime},R_2^{\prime}\right)$ are any two rate-pairs in ${\mathcal R}_{\rm w/o-TS}^{\rm S}$, and achievable by the jamming beamforming vectors ${\bf w}_1$ and ${\bf w}_1^{\prime}$, respectively.

Notice that any point on the segment connecting $\left(R_1,R_2\right)$ and $\left(R_1^{\prime},R_2^{\prime}\right)$  is achievable by time-sharing based jamming strategy. This is because, if the legitimate monitor uses the jamming beamforming vector ${\bf w}_1$ during $0\leq \tau \leq 1$ portion of each block time and the jamming beamforming vector ${\bf w}_1^{\prime}$ during the rest $1-\tau$ portion of each block time, then the achievable rate-pair
$\left(\tau R_1\!+\!\left(1\!-\!\tau\right)R_1^{\prime},\tau R_2\!+\!\left(1\!-\!\tau\right)R_2^{\prime}\right)$ is achievable. Therefore, by adjusting the value of $\tau$, any point on the segment connecting these two points is achievable.

Note that the region ${\mathcal R}_{\rm w/-TS}^{\rm S}$ is essentially obtained from the convex hull operation over the region ${\mathcal R}_{\rm w/o-TS}^{\rm S}$, thus we have  ${\mathcal R}_{\rm w/o-TS}^{\rm S}\subseteq {\mathcal R}_{\rm w/-TS}^{\rm S}$, where the equality holds if and only if ${\mathcal R}_{\rm w/o-TS}^{\rm S}$ is a convex set. In other words, when the region ${\mathcal R}_{\rm w/o-TS}^{\rm S}$ is non-convex, the achievable rate region for the suspicious links can be further enlarged by time-sharing based jamming strategy.

Then, similar to \eqref{E9sss11}, we define the achievable eavesdropping rate region for the MMSE receiver case with time-sharing based jamming strategy as the intersection of the regions ${\mathcal R}_{\rm MMSE}^{\rm E}$ and ${\mathcal R}_{\rm w/-TS}^{\rm S}$, i.e.,
\begin{equation}\label{E9sss1ss1}
{\mathcal R}_{\rm w/-TS}^{\rm MMSE}=  {\mathcal R}_{\rm MMSE}^{\rm E} \cap { {\mathcal R}}_{\rm w/-TS}^{\rm S}.
\end{equation}

It is easy to verify that ${\mathcal R}_{\rm w/o-TS}^{\rm MMSE}\subseteq {\mathcal R}_{\rm w/-TS}^{\rm MMSE}$ since ${\mathcal R}_{\rm w/o-TS}^{\rm S} \subseteq {\mathcal R}_{\rm w/-TS}^{\rm S}$. Therefore, the time-sharing based jamming strategy has the potential to enlarge the achievable eavesdropping rate region in general.

\section{Extension to MMSE-SIC Receiver Case}

\begin{figure}[!ht]
\centering
\includegraphics[scale=1]{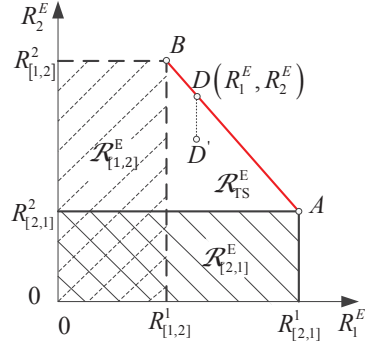}
\caption{An illustrative example of the achievable rate region ${\mathcal R}_{\rm MMSE-SIC}^{\rm E}$ for the two eavesdropping links at the monitor. }\label{201s80725wa1} 
\end{figure}
In this section, we extend our study for characterizing the achievable eavesdropping rate region from the linear MMSE receiver in Section \uppercase\expandafter{\romannumeral2} to the non-linear MMSE-SIC receiver. To be specific, we first describe the achievable rate region for the eavesdropping links with the MMSE-SIC receiver, which is always larger than that with the MMSE receiver. Then, we characterize the achievable eavesdropping rate region for the MMSE-SIC receiver and analyze the required scheme to achieve any point within this region.

The fundamental principle of MMSE-SIC is for the monitor to decode one suspicious message with MMSE receiver first, and then subtract it from the received signal before decoding the other suspicious message \cite{SIC}.
Let ${\boldsymbol \pi}=\left[\pi_1,\pi_2\right]$ denote the decoding order vector: the message of the $\pi_1$-th suspicious link is decoded first, and its effect is then removed from the received signal. Consequently, the achievable rates of the $\pi_1$-th and $\pi_2$-th eavesdropping links are respectively  given by
\begin{equation}\label{E311aaaa}
{R}_{\left[\pi_1,\pi_2\right]}^{\pi_1}={\rm log}\left(1+P_{\pi_1}{\bf h}_{\pi_1,m}^H\left(P_{\pi_2}{\bf h}_{{\pi_2},m}{\bf h}_{{\pi_2},m}^H+\sigma_m^2{\bf I}\right)^{-1}{\bf h}_{\pi_1,m}\right),
\end{equation}
\begin{equation}\label{E3122}
{R}_{\left[\pi_1,\pi_2\right]}^{\pi_2}={\rm log}\left(1+\frac{P_{\pi_2}{\bf h}_{\pi_2,m}^H{\bf h}_{\pi_2,m}}{\sigma_m^2}\right).
\end{equation}

For example, if the legitimate monitor decodes the message of the first suspicious link first, i.e., $\pi_1=1$ and $\pi_2=2$, then the achievable rates of the first  and second eavesdropping links are explicitly written as
$R_{\left[1,2\right]}^{1}={\rm log}\left(1+P_{1}{\bf h}_{1,m}^H\left({P_2{\bf h}_{2,m}{\bf h}_{2,m}^H}+\sigma_m^2{\bf I}\right)^{-1}{\bf h}_{1,m}\right)$ and $R_{\left[1,2\right]}^{2}={\rm log}\left(1+\frac{P_{2}{\bf h}_{2,m}^H{\bf h}_{2,m}}{\sigma_m^2}\right)$, respectively. Notice that thanks to the cancellation of interference from Alice 1 for Bob 2's reception, we have $R_{\left[1,2\right]}^{2}> R_{\left[2,1\right]}^{2}$, as shown in Fig. \ref{201s80725wa1}. Similarly, we also have $R_{\left[2,1\right]}^{1}> R_{\left[1,2\right]}^{1}$.

Let ${\mathcal R}_{\left[1,2\right]}^{\rm E}$ and ${\mathcal R}_{\left[2,1\right]}^{\rm E}$ denote the achievable rate regions for the suspicious links associated with decoding orders ${\boldsymbol \pi}=\left[1,2\right]$ and ${\boldsymbol \pi}=\left[2,1\right]$, respectively, which are two rectangular regions and given by (as shown in Fig. \ref{201s80725wa1})
\begin{equation}\label{E9sss20}
{\mathcal R}_{\left[1,2\right]}^{\rm E}=\Bigg\{\left(R_1^E,R_2^E\right)\Big| 0\leq R_1^E \leq R_{\left[1,2\right]}^1; 0\leq R_2^E \leq R_{\left[1,2\right]}^2\Bigg\},
\end{equation}
\begin{equation}\label{E9sss21}
{\mathcal R}_{\left[2,1\right]}^{\rm E}=\Bigg\{\left(R_1^E,R_2^E\right)\Big| 0\leq R_1^E \leq R_{\left[2,1\right]}^1; 0\leq R_2^E \leq R_{\left[2,1\right]}^2\Bigg\}.
\end{equation}

Note that the regions  ${\mathcal R}_{\left[1,2\right]}^{\rm E}$ and ${\mathcal R}_{\left[2,1\right]}^{\rm E}$  are achievable with the decoding orders ${\boldsymbol \pi}=\left[1,2\right]$ and ${\boldsymbol \pi}=\left[2,1\right]$, respectively.
By allowing the legitimate monitor to arbitrarily choose the decoding order and employ time-sharing between these two decoding orders, the achievable region for the eavesdropping links can be characterized by (as shown in Fig. \ref{201s80725wa1})
\begin{equation}\label{E9ss}
{\mathcal R}_{\rm MMSE-SIC}^{\rm E}={\rm Conv} \left({\mathcal R}_{\left[1,2\right]}^{\rm E}\bigcup {\mathcal R}_{\left[2,1\right]}^{\rm E}\right),
\end{equation}
where ${\rm Conv}\left(\mathcal {S}\right)$ denotes the convex hull of the set $\mathcal {S}$.

The region  ${\mathcal R}_{\rm TS}^{\rm E}$ in Fig. \ref{201s80725wa1}  is added thanks to the operation of time-sharing between different decoding orders. $A=\left(R_{\left[2,1\right]}^1,R_{\left[2,1\right]}^2\right)$ and $B=\left(R_{\left[1,2\right]}^1,R_{\left[1,2\right]}^2\right)$  are two corner points corresponding the decoding orders ${\boldsymbol \pi}=\left[2,1\right]$ and ${\boldsymbol \pi}=\left[1,2\right]$, respectively. Any point $D=\left(R_1^E,R_2^E\right)$, with $R_1^E \in \left[R_{\left[1,2\right]}^1,R_{\left[2,1\right]}^1\right]$ and $R_2^E \in \left[R_{\left[2,1\right]}^2,R_{\left[1,2\right]}^2\right]$, on the line segment connecting the points $A$ and $B$ is achievable by employing the time-sharing between the two decoding orders, with the following adjustable time-sharing factor:
\begin{equation}\label{E9ss1s}
\varsigma\left(R_1^E\right)=\frac{R_{\left[2,1\right]}^1-R_1^E}{R_{\left[2,1\right]}^1-R_{\left[1,2\right]}^1}.
\end{equation}

That is,  the legitimate monitor uses the decoding order ${\boldsymbol \pi}=\left[1,2\right]$ during the first $\varsigma\left(R_1^E\right)$ proportion of each time slot and the other decoding order ${\boldsymbol \pi}=\left[2,1\right]$ during the rest $1-\varsigma\left(R_1^E\right)$ proportion of each time slot. Specifically, $A$ and $B$ can be regarded as two specifical points corresponding the two time-sharing factors $\varsigma\left(R_{\left[2,1\right]}^1\right)=0$ and $\varsigma\left(R_{\left[1,2\right]}^1\right)=1$, respectively. Accordingly, the triangular region ${\mathcal R}_{\rm TS}^{\rm E}$ below the line segment $\overline{AB}$ is achievable by using the time-sharing between different decoding orders. In Fig. \ref{201s80725wa1}, any
interior  point $D^{'}$ in ${\mathcal R}_{\rm TS}^{\rm E}$ is also achievable with the same time-sharing factor as $\varsigma\left(R_1^E\right)$.

Then, similar to the definition of ${\mathcal R}_{\rm w/o-TS}^{\rm MMSE}$ in  \eqref{E9sss11}, we define the achievable eavesdropping rate region for the MMSE-SIC receiver case without time-sharing among different jamming transmit covariance matrices as\footnote{Note that similar to the definition of ${\mathcal R}_{\rm w/-TS}^{\rm MMSE}$ in  \eqref{E9sss1ss1}, we can also define the achievable eavesdropping rate region for the MMSE-SIC receiver case with time-sharing among different jamming transmit covariance matrices as ${\mathcal R}_{\rm w/-TS}^{\rm MMSE-SIC}=  {\mathcal R}_{\rm MMSE-SIC}^{\rm E} \cap { {\mathcal R}}_{\rm w/-TS}^{\rm S}$.  Since ${\mathcal R}_{\rm w/o-TS}^{\rm S} \subseteq {\mathcal R}_{\rm w/-TS}^{\rm S}$, we have ${\mathcal R}_{\rm w/o-TS}^{\rm MMSE-SIC}\subseteq {\mathcal R}_{\rm w/-TS}^{\rm MMSE-SIC}$. }
\begin{equation}\label{E9sssss}
{\mathcal R}_{\rm w/o-TS}^{\rm MMSE-SIC}=  {\mathcal R}_{\rm MMSE-SIC}^{\rm E} \cap { {\mathcal R}}_{\rm w/o-TS}^{\rm S}.
\end{equation}

 \begin{figure}[!ht]
\centering
\includegraphics[scale=1]{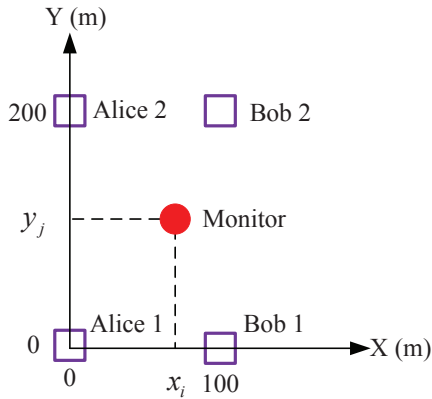}
\caption{System setup for simulation. }\label{201s80725was} 
\end{figure}
Note that since the MMSE-SIC receiver yields larger achievable rate region for the eavesdropping links than that by the MMSE receiver, i.e.,
${\mathcal R}_{\rm MMSE}^{\rm E}\subseteq {\mathcal R}_{\rm MMSE-SIC}^{\rm E}$, we have ${\mathcal R}_{\rm w/o-TS}^{\rm MMSE} \subseteq {\mathcal R}_{\rm w/o-TS}^{\rm MMSE-SIC}$.

Finally, we discuss the scheme for the legitimate monitor to achieve any point in ${\mathcal R}_{\rm w/o-TS}^{\rm MMSE-SIC}$, i.e., $\forall z=\left(R_1^z,R_2^z\right) \in {\mathcal R}_{\rm w/o-TS}^{\rm MMSE-SIC}$.  It is necessary for the legitimate monitor to jointly choose the time-sharing factor between different decoding orders as well as the jamming beamforming vector so as to guarantee $z \in {\mathcal R}_{\rm MMSE-SIC}^{\rm E}$ and $z \in {\mathcal R}_{\rm w/o-TS}^{\rm S}$ concurrently. While the jamming beamforming is already analyzed and given in Section IV, we only need to update the time-sharing factor of decoding orders for ${\mathcal R}_{\rm MMSE-SIC}^{\rm E}$ region only, as given by
\begin{eqnarray}\label{20181003ds} \varsigma^*=
\begin{cases}
0, &{\rm if~}
z \in{\mathcal R}_{\left[1,2\right]}^{\rm E}\cr \varsigma\left(R_1^z\right) ~{\rm in~ \eqref{E9ss1s}}, &{\rm if~} z  \in{\mathcal R}_{\rm TS}^{\rm E}\cr 1, &{\rm if~}
z \in{\mathcal R}_{\left[2,1\right]}^{\rm E}\end{cases}.
\end{eqnarray}

 \section{Numerical Results}

In this section, we present numerical examples to show the achievable eavesdropping rate regions by our proactive jamming approach. We consider the two-dimensional (2D) Cartesian coordinate system with X and Y axes on the ground plane shown in Fig. \ref{201s80725was}, where Alice 1, Alice 2, Bob 1, and Bob 2 are respectively located at  $\left(0,0\right)$, $\left(0, 200~{\rm meters}\right)$, $\left(100~{\rm meters},0\right)$, and $\left(100~{\rm meters},200~{\rm meters}\right)$, and  $M=\left(x_i, y_j\right)$ denotes the location of the legitimate monitor. We adopt the distance-dependent pass loss model, which is given by
\begin{figure}[!ht]
\centering
\includegraphics[scale=0.8]{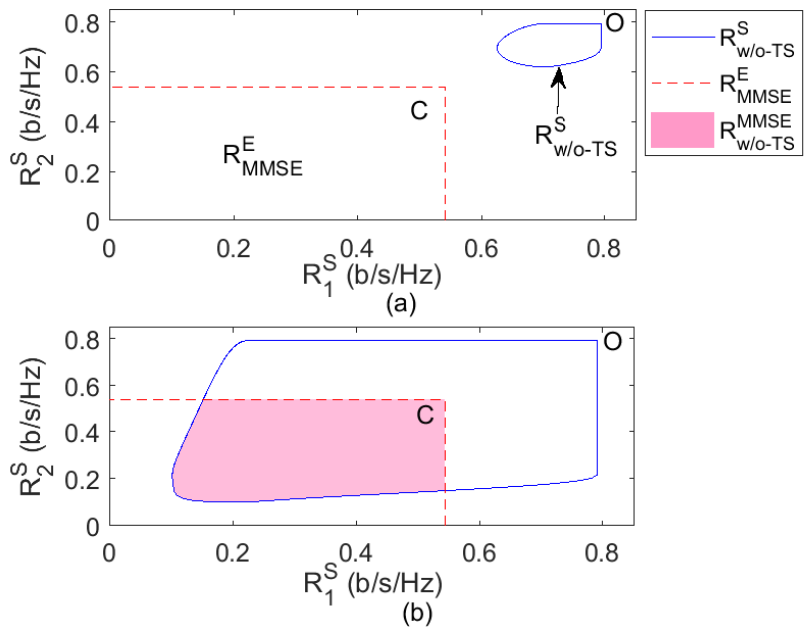}
\caption{The achievable eavesdropping rate regions for the case of MMSE receiver with different jamming transmit power budgets at the legitimate monitor. (a) $P_{\rm max}=4$ dBm. (b) $P_{\rm max}=20$ dBm. }\label{fig:subfig} 
\end{figure}
\begin{equation} \label{ChannelModel}
L = A_0\left(\frac{d}{d_0}\right)^{-\alpha},
\end{equation}
where $A_0 = 10^{-3}$, $d$ denotes the distance between the transmitter and the receiver, $d_0=1$~meter is a reference distance, and $\alpha=2.5$ is the path loss exponent. All the channels  are randomly generated from independently and identically
distributed (i.i.d) Rayleigh fading with zero mean and variance specified by \eqref{ChannelModel}. Unless otherwise stated, the numbers of transmitting and receiving antennas at the legitimate monitor are set to $N_t=N_r=2$. We set the maximum transmit power at the suspicious transmitters as $P_1=P_2= 10$ dBm, and  the noise power at each receiver as $\sigma^2=-70$ dBm. All the results are obtained by averaging over 10,000 independent channel realizations.

\subsection{Achievable Eavesdropping Rate Region with MMSE Receiver. }

First, we examine  the achievable eavesdropping rate regions for the case of MMSE receiver with different jamming transmit power budgets at the legitimate monitor, i.e., $P_{\rm max}=4$ dBm or $P_{\rm max}=20$ dBm, where the legitimate monitor is located at $M=\left(100~{\rm meters},100~{\rm meters}\right)$. In Fig. \ref{fig:subfig} the rectangle area with vertex $C$  denotes the achievable rate region for the eavesdropping links ${\mathcal R}_{\rm MMSE}^{\rm E}$. The point $O$ corresponds to the achievable rate pair of passive eavesdropping without jamming. In Fig. \ref{fig:subfig}, the point $O$ is outside the region $R_{\rm MMSE}^{\rm E}$, which tells that the passive eavesdropping cannot help the legitimate monitor eavesdrop any suspicious link. Furthermore, the irregular area with vertex $O$ represents the achievable rate region for the suspicious links ${\mathcal R}_{\rm w/o-TS}^{\rm S}$, achieved by the jamming-assisted proactive eavesdropping approach.
From Fig. \ref{fig:subfig}, we can see that with larger jamming transmit power budget, the legitimate monitor can use jamming to further decrease the rates of suspicious links, allowing the legitimate monitor to decode both  suspicious links' data successfully. For example, the achievable rate region of the suspicious links ${\mathcal R}_{\rm w/o-TS}^{\rm S}$  in Fig. \ref{fig:subfig}(a) is much smaller than that in Fig. \ref{fig:subfig}(b). Specifically, the intersection of the region ${\mathcal R}_{\rm w/o-TS}^{\rm S}$ with the achievable rate region for the eavesdropping links $R_{\rm MMSE}^{\rm E}$  is an empty set in Fig. \ref{fig:subfig}(a), and thus the legitimate monitor cannot eavesdrop any suspicious link; whereas the intersection of the regions ${\mathcal R}_{\rm w/o-TS}^{\rm S}$ and $R_{\rm MMSE}^{\rm E}$ is ${\mathcal R}_{\rm w/o-TS}^{\rm MMSE}$ thanks to the stronger jamming, shown as the shaded area in Fig. \ref{fig:subfig}(b).
\begin{figure}
  \centering
  \subfigure[Legitimate monitor's location at $M=\left(50~{\rm meters},70~{\rm meters}\right)$.]{
    \includegraphics[scale=0.59]{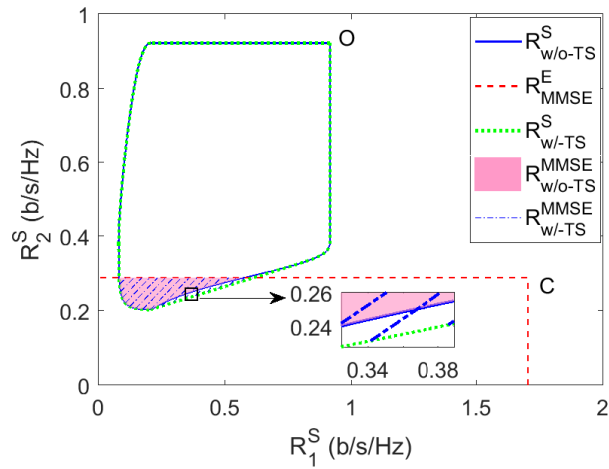}
  }
  \subfigure[Legitimate monitor's location at $M=\left(50~{\rm meters},130~{\rm meters}\right)$.]{
    \includegraphics[scale=0.59]{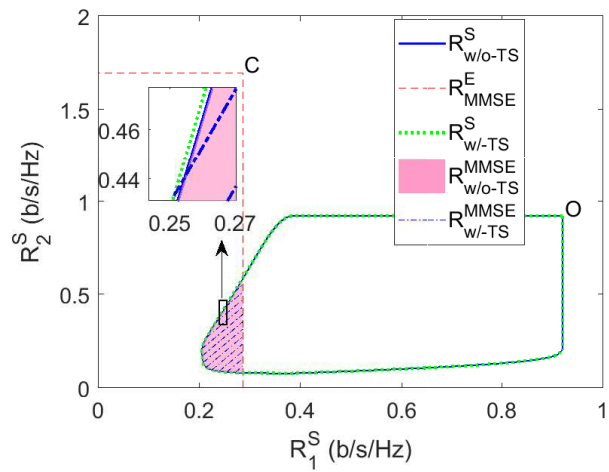}
  }
  \caption{The achievable eavesdropping rate regions with different locations of the legitimate monitor.}
  \label{fig:subfig1} 
\end{figure}

Next, we study the impact of the legitimate monitor's location on the achievable eavesdropping rate region  with $M=\left(50~{\rm meters},70~{\rm meters}\right)$ and $M=\left(50~{\rm meters},130~{\rm meters}\right)$, respectively.
The jamming transmit power is set to be $P_{\rm max}=20$ dBm. From Fig. \ref{fig:subfig1}(a), we can see that the achievable rate region for the eavesdropping links $R_{\rm MMSE}^{\rm E}$ is a wide rectangle area, i.e., $R_1^E> R_2^E$. This is due to the fact that when $M=\left(50~{\rm meters},70~{\rm meters}\right)$, the legitimate monitor is closer to Alice 1 than to Alice 2 so that the first eavesdropping link experiences the minimal distance-dependent
signal attenuation and co-channel interference caused by the second suspicious transmitter. From Fig. \ref{fig:subfig1}(a), we can also observe that the legitimate monitor can eavesdrop the first suspicious link successfully even without sending any jamming signals, i.e., $R_1^E > R_1^S\left({\bf 0}\right)$, yet cannot overhear from the second suspicious link. Our jamming-assisted approach targets at the rate reduction of the second suspicious link, helping the legitimate monitor overhear both the suspicious links.  On the other hand, when the legitimate monitor is located at $M=\left(50~{\rm meters},130~{\rm meters}\right)$, it is closer to Alice 2  than to Alice 1. Thus, from Fig. \ref{fig:subfig1}(b), we can see that in this case $R_{\rm MMSE}^{\rm E}$ is a tall rectangle area (i.e., $R_1^E< R_2^E$) and the legitimate monitor can eavesdrop the second suspicious link successfully without sending any jamming signals (i.e., $R_2^E > R_2^S\left({\bf 0}\right)$), yet cannot overhear from the first suspicious link. Our jamming-assisted approach aims at reducing the first suspicious link's rate, and thus helps the legitimate monitor overhear both the suspicious links successfully.
Furthermore, it can be observed from Figs. \ref{fig:subfig1}(a) and \ref{fig:subfig1}(b) that  time-sharing between different jamming beamforming vectors can enlarge the achievable rate region of the suspicious links over the case without time sharing, i.e., $R_{\rm w/o-TS}^{\rm S}\subset R_{\rm w/-TS}^{\rm S}$, which further results in a larger achievable eavesdropping rate region, i.e., $R_{\rm w/o-TS}^{\rm MMSE} \subset R_{\rm w/-TS}^{\rm MMSE}$.

\begin{figure}[!ht]
\centering
\includegraphics[scale=0.8]{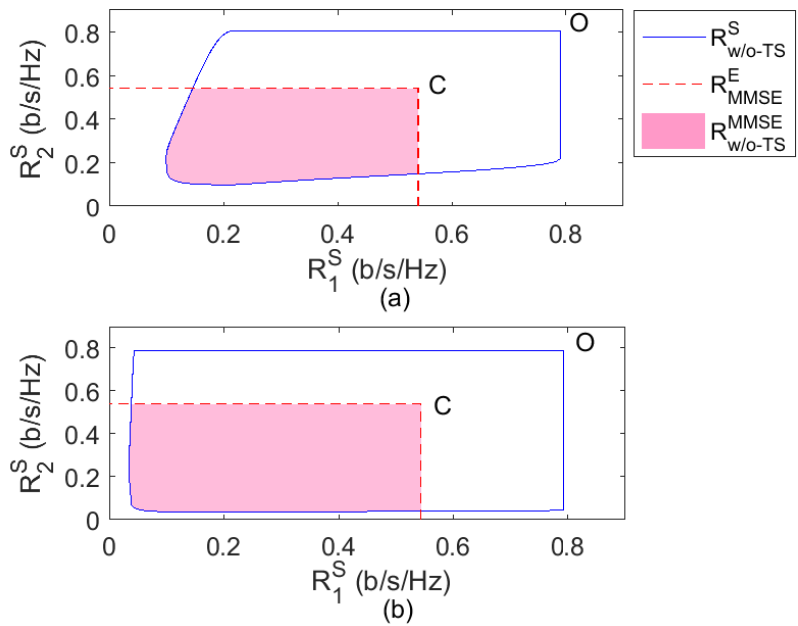}
\caption{The achievable eavesdropping rate region versus $N_t$. (a) $N_t=2$. (b) $N_t=5$. }\label{20190715ss} 
\end{figure}

Furthermore, we show the impact of the number of jamming antennas $N_t$ on the achievable eavesdropping rate region with $N_t=2$ and $N_t=5$, respectively.  The legitimate monitor is locatd  at $M=\left(100~{\rm meters},100~{\rm meters}\right)$, and the jamming transmit power budget is set as $P_{\rm max}=20$ dBm. Comparing Figs. \ref{20190715ss}(a) and \ref{20190715ss}(b), we can clearly see that the achievable eavesdropping rate region is significantly enlarged as the number of jamming antennas $N_t$ increases from 2 to 5. This is because, as $N_t$ increases, the legitimate monitor has more degrees of freedom in spatial domain to design jamming transmit covariance matrix, which enhances the jamming performance and  thus enlarges the
achievable rate region of suspicious links $R_{\rm w/o-TS}^{\rm S}$.

\subsection{Achievable Eavesdropping Rate Region with MMSE-SIC Receiver }
\begin{figure}
  \centering
  \subfigure[Legitimate monitor's location at $M=\left(100~{\rm meters},100~{\rm meters}\right)$.]{
    \includegraphics[scale=0.6]{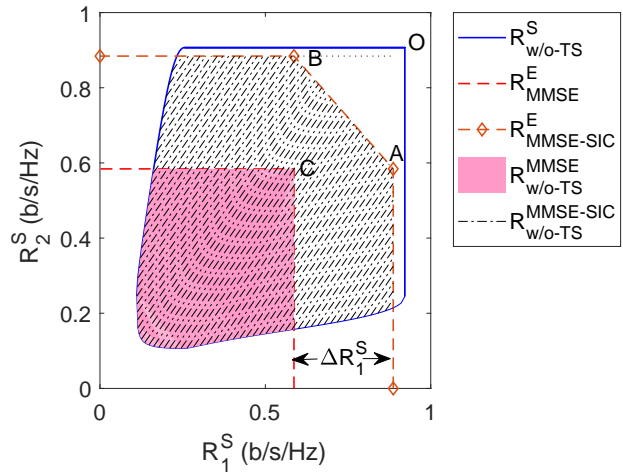}
  }
  \subfigure[Legitimate monitor's location at $M=\left(0,100~{\rm meters}\right)$.]{
    \includegraphics[scale=0.6]{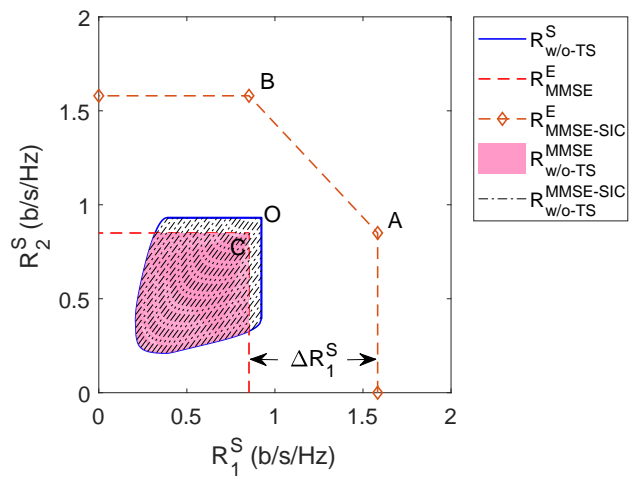}
  }
  \caption{The achievable eavesdropping rate regions for both the cases of MMSE and MMSE-SIC receiver under different locations of the legitimate monitor.}
  \label{fig:subfigd1} 
\end{figure}
In Fig. \ref{fig:subfigd1}, we compare the achievable eavesdropping rate regions for both MMSE and MMSE-SIC receiver cases under different locations
of the legitimate monitor, i.e., $M=\big(100~{\rm meters},\\100~{\rm meters}\big)$ and $M=\left(0,100~{\rm meters}\right)$, respectively. The monitor's  jamming transmit power is set to $P_{\rm max}=20$ dBm. As we can see from
Fig. \ref{fig:subfigd1}, the achievable eavesdropping rate region for the MMSE-SIC receiver case $R_{\rm w/o-TS}^{\rm MMSE-SIC}$ is significantly larger than that for the MMSE receiver case $R_{\rm w/o-TS}^{\rm MMSE}$, i.e., $R_{\rm w/o-TS}^{\rm MMSE}\subset R_{\rm w/o-TS}^{\rm MMSE-SIC}$. This is because of the joint operations of SIC and time-sharing between two decoding orders at the legitimate receiver for improving both the eavesdropping links. Furthermore, comparing Figs. \ref{fig:subfigd1}(a) and \ref{fig:subfigd1}(b), we can observe that the achievable eavesdropping rate regions for the MMSE receiver case are enlarged by moving the location of the legitimate monitor from $M=\left(100~{\rm meters},100~{\rm meters}\right)$ to $M=\left(0,100~{\rm meters}\right)$. Meanwhile, during this process, the difference between the achievable rate regions $R_{\rm MMSE}^{\rm E}$ and $R_{\rm MMSE-SIC}^{\rm E}$ becomes large, i.e., the value of $\Delta R_1^S$ increases approximately  from 0.35 bits per second per hertz (b/s/Hz) to 0.85 b/s/Hz. This is due to the fact that when $M=\left(0,100~{\rm meters}\right)$, the legitimate monitor is in a strong co-channel interference environment from both Alice 1 and Alice 2, and thus the performance advantage of the MMSE-SIC receiver is more significant.

\subsection{Achievable Eavesdropping Rate Region with Imperfect Self-Interference Cancellation}
In this subsection, we provide numerical results to show the impact of imperfect self-interference cancellation on the eavesdropping performance of our proposed solutions. Particularly, we focus on evaluating the effect of imperfect self-interference cancellation on the achievable eavesdropping rate region of the MMSE receiver case.

 To begin, for the case with imperfect self-interference cancellation, the achievable rate of the $i$th eavesdropping link is modified from \eqref{E311} as
$R_i^E\left({\bf Q}\right)={\rm log}\big(1+\frac{P_i{\bf h}_{i,m}^H{\bf h}_{i,m}}{\rho{\rm Tr}\left({\bf H}_{ee}{\bf Q}{\bf H}_{ee}^H\right)+{P_j{\bf h}_{j,m}{\bf h}_{j,m}^H}+\sigma_m^2{\bf I}}\big), \forall i,j\in\{1,2\}, i\neq j,$
where ${\bf H}_{ee}\in {\mathbb{C}^{N_r\times N_t }}$ denotes the feedback loop channel between the transmitter antennas and the receiver antennas of the legitimate monitor, and $\rho \in \left[0,1\right]$ is the self-interference cancellation coefficient, which represents the degree of passive self-interference suppression, i.e., $\rho=0$ means perfect self-interference cancellation. Let $\mathcal {A}$ denote the set of all jamming covariance matrices ${\bf Q}$ required to achieve the region ${\mathcal R}_{\rm w/o-TS}^{\rm S}$ defined in \eqref{E9s1}. Then, the minimum achievable rate rate of the $i$th eavesdropping link with imperfect self-interference cancellation, denoted by $R_i^{SI}$, is given by
$R_i^{SI}=\min_{\forall {\bf Q} \in \mathcal {A}} R_i^E\left({\bf Q}\right), \forall i\in\{1,2\}$. By replacing  $R_i^{E}$ in \eqref{E9s} with $R_i^{SI}$ for $\forall i\in\{1,2\}$, we obtain an achievable (lower-bound) rate region for the eavesdropping links with imperfect self-interference cancellation and denote it as ${\mathcal R}_{\rm MMSE}^{\rm E-SI}$. Then, we can further  obtain the achievable eavesdropping rate region with imperfect self-interference cancellation, denoted by ${\mathcal R}_{\rm w/o-TS}^{\rm MMSE-SI}$,  by replacing ${\mathcal R}_{\rm MMSE}^{\rm E}$ in \eqref{E9sss11} with ${\mathcal R}_{\rm MMSE}^{\rm E-SI}$.

For comparison, we also consider a heuristic  null space-based jamming beamforming design and show the corresponding achievable eavesdropping rate region, denoted by ${\mathcal R}_{\rm w/o-TS}^{\rm MMSE-NS}$. In this design, the jamming covariance
matrix is chosen such that ${\bf H}_{ee}{\bf Q} {\bf H}_{ee}^H= {\bf 0}$, i.e., jamming signals are sent in the null space of the feedback loop channel matrix ${\bf H}_{ee}$,  thus leading to no self-interference regardless of the value of $\rho$. Let ${\bf V}\in {\mathbb{C}^{N_t\times \left(N_t- r\right)}}$ denote the orthogonal basis of the null space of ${\bf H}_{ee}$, with $r$ being the rank of ${\bf H}_{ee}$.
 \begin{figure}[!ht]
\centering
\includegraphics[scale=0.8]{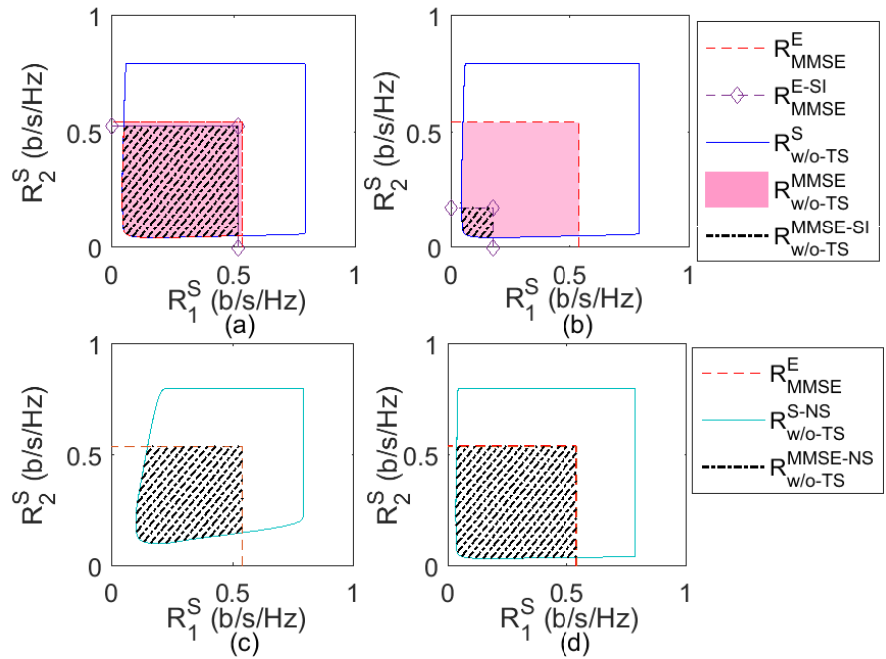}
\caption{Comparison of achievable eavesdropping rate regions of our proposed solutions with
and without perfect self-imperfect cancellation. (a) $N_t=5$, $\rho = -75$ dB; (b) $N_t=5$, $\rho = -55$ dB; (c) $N_t=5$, $\rho = -55$ dB; (d) $N_t=7$, $\rho = -55$ dB. }\label{20190715s} 
\end{figure}
Then, the jamming covariance matrix ${\bf Q}$ can be expressed as ${\bf Q}={\bf V}{\bf {\bar Q}}{\bf V}^H$, where ${\bf {\bar Q}}$ is a $\left(N_t- r\right) \times \left(N_t- r\right)$ positive semi-definite matrix.
By substituting ${\bf Q}={\bf V}{\bf {\bar Q}}{\bf V}^H$ into \eqref{R1} and after defining ${\bf {\tilde{g}}}_i={\bf V}^H{\bf g}_i$, we can use the same approach as in previous Section \Rmnum {3} to design ${\bf {\bar Q}}$ and then characterize the achievable region ${\mathcal R}_{\rm w/o-TS}^{\rm MMSE-NS}$. Moreover, we note that the orthogonal basis ${\bf V}$ can be determined from the singular value decomposition of ${\bf H}_{ee}$  (following the same approach in \cite{Yong2}). It is also worth pointing out that the above null space-based jamming design works only when $N_t >r$. Obviously, since $r\leq {\rm min} \left\{N_t,N_r\right\}$, a sufficient condition for $N_t >r$ to hold is that $N_t >N_r$, i.e., the legitimate monitor allocates more antennas at its transmitter than that at its receiver.



In Figs. \ref{20190715s}(a) and \ref{20190715s}(b), we show the impact of self-interference on our proposed solution under different self-interference cancellation coefficients, i.e., $\rho = -75$ dB and $\rho = -55$ dB, respectively, while in Figs. \ref{20190715s}(c) and \ref{20190715s}(d), we plot the achievable eavesdropping rate region of the null space-based jamming design with $N_t=5$ and $N_t=7$, respectively. The legitimate monitor is location  at $M=\left(100~{\rm meters},100~{\rm meters}\right)$, and the jamming transmit power budget is set to be $P_{\rm max}=20$ dBm. From Fig. \ref{20190715s}(a), we can clearly see that the region ${\mathcal R}_{\rm w/o-TS}^{\rm MMSE-SI}$ is just slightly smaller than the region ${\mathcal R}_{\rm w/o-TS}^{\rm MMSE}$ (obtained by perfect self-interference cancellation). This observation suggests that ignoring the self-interference will not significantly degrade the eavesdropping performance if the self-interference cancellation coefficient is small enough, i.e. $\rho \leq -75$ dB. Please note that the jointly analog and digital self-interference cancellation technique (recently reported in \cite{SIC110}) is able to achieve up to 110 dB self-interference reduction, i.e., $\rho = -110$ dB. Thus, it is reasonable to ignore the self-interference when legitimate monitor adopts this advanced self-interference cancellation technique. Of course, if $\rho$ increases to a huge level, e.g., $-55$ dB in Fig. 11(b), then the region ${\mathcal R}_{\rm w/o-TS}^{\rm MMSE-SI}$  is significantly smaller than the region ${\mathcal R}_{\rm w/o-TS}^{\rm MMSE}$. In this case, the legitimate monitor may choose the null space-based jamming strategy to send jamming signals in order to obtain better eavesdropping performance. This is because, from Figs. \ref{20190715s}(b) and \ref{20190715s}(c), it can be observed that the null space-based jamming design achieves better eavesdropping rate region than our proposed solution with imperfect self-interference cancellation. Furthermore, comparing Figs. \ref{20190715s}(c) and \ref{20190715s}(d), we observe that the achievable eavesdropping rate region of the null space-based jamming design is significantly enlarged as the number of jamming antennas $N_t$ increases from 5 to 7. This is due to the fact that, as $N_t$ increases, the legitimate monitor has more degrees of freedom in the null space of ${\bf H}_{ee}$
to design jamming transmit covariance matrix, thus making it more capable of changing the achievable rate of suspicious links (i.e., increasing the corresponding achievable rate region of suspicious links, denoted by ${\mathcal R}_{\rm w/o-TS}^{\rm S-NS}$).

\section{Conclusion}
In this paper, we considered the proactive eavesdropping over two suspicious communication links scenario. Specifically, we first characterized the achievable eavesdropping rate region for the MMSE receiver case with or without time sharing the jamming transmit covariance matrix. For this purpose, we derived the closed-form expressions of the upper and lower boundary points of the achievable rate region for the two suspicious links and analyzed the monotonicity of the upper and lower boundary curves. Furthermore, we extended our study to the MMSE-SIC receiver case and characterized the corresponding achievable eavesdropping rate region, by jointly optimizing the time-sharing factor. Simulation results showed that significant performance gain is achieved with the proactive eavesdropping over the passive eavesdropping, and the achievable eavesdropping rate region for the MMSE-SIC receiver case is notably larger than that for the MMSE receiver case, especially when the legitimate monitor is in a strong co-channel interference environment.

There are several directions to study beyond this work in the future. For example, how to extend the model of two suspicious communication links to multiple links is an unsolved problem, while considering the more advanced transmitters/receivers of the suspicious links (e.g., with multiple antennas) will make the problem more challenging.  For example, if
the suspicious users are able to detect the jamming attack, they may adopt anti-jamming methods such as random frequency hopping to avoid jamming. In this case, how to study the interplay between the legitimate monitor and suspicious users is an interesting open problem.


\appendices
\section{Proof of Lemma \ref{P20180903a}}
Let ${\bf g}_2^{\perp}$ denote the projection of ${\bf g}_1$ into the null space of ${\bf \hat g}_2$, i.e., ${\bf g}_2^{\perp}=\left({\bf I}-{\bf \hat g}_2{\bf \hat g}_2^H\right){\bf g}_1$.
Thus, ${\bf \hat g}_2^{\perp}$ and ${\bf \hat g}_2$ are two orthonormal vectors, where  ${\bf \hat g}_2^{\perp}=\frac{{\bf g}_2^{\perp}}{\left\|{\bf g}_2^{\perp}\right\|}$.  Then, we can define a matrix ${\boldsymbol \Pi}=\left[{\bf \hat g}_2~{\bf \hat g}_2^{\perp}~{\boldsymbol \Pi}_\bot\right]$, with ${\boldsymbol \Pi}_\bot \in {\mathbb{C}^{N_t \times \left(N_t-2\right)}}$ being the orthogonal complement of $\left[{\bf \hat g}_2~{\bf \hat g}_2^{\perp}\right]$. Thus,
${\boldsymbol \Pi}$ is a unitary matrix, i.e., ${\boldsymbol \Pi}^H{\boldsymbol \Pi}={\bf I}$, with columns spanning the $N_t$-dimensional space. Therefore, any jamming beamforming vector ${\bf w}$ can be expressed as
\begin{equation}\label{E17s23s}
{\bf w}=\left[{\bf \hat g}_2~{\bf \hat g}_2^{\perp}~{\boldsymbol \Pi}_\bot\right]
\left[ {\begin{array}{*{20}{c}}
\alpha\\
\beta\\
{\bf z}
\end{array}} \right],
\end{equation}
where $\alpha \in {\mathbb{C}}$, $\beta \in {\mathbb{C}}$, and ${\bf z} \in {\mathbb{C}^{\left(N_t-2\right) \times 1}}$.

Using the fact that ${\bf g}_1^H{\boldsymbol \Pi}_\bot$=${\bf g}_2^H{\boldsymbol \Pi}_\bot=0$, and ${\bf g}_2^H{\bf \hat g}_2^{\perp}=0$, we have
\begin{equation}\label{20180903d}
{\bf g}_1^H{\bf w}=\alpha{\bf g}_1^H{\bf \hat g}_2+\beta{\bf g}_1^H{\bf \hat g}_2^{\perp},
\end{equation}
\begin{equation}\label{20180903e}
{\bf g}_2^H{\bf w}=\alpha\left\|{\bf g}_2\right\|.
\end{equation}

By substituting \eqref{20180903d} and \eqref{20180903e} into problem \eqref{20180903c}, we can express it alternatively as
\begin{equation}\label{20180903f}
\begin{split}
&\min_{ \alpha,\beta,{\bf z}}~~\left|\alpha\right|^2\left\|{\bf g}_2\right\|^2\\
&~~s.t.~~\left|\alpha{\bf g}_1^H{\bf \hat g}_2+\beta{\bf g}_1^H{\bf \hat g}_2^{\perp}\right|^2=\phi\left(R_1\right)\\
&~~~~~~~~\left|\alpha\right|^2+\left|\beta\right|^2+ \left\|\bf z\right\|^2\leq P_{\rm max}
\end{split}
\end{equation}

It is obvious that $\bf z$ has no influence on both the objective function and  the first constraint in problem \eqref{20180903f}, but only consumes power budget $P_{\rm max}$. Therefore, $\bf z$ should be set to zero so as to enlarge the feasible set. Thus, the optimal jamming vector ${\bf w}$ has the form of ${\bf w}=\alpha{\bf \hat g}_2+\beta {\bf \hat g}_2^{\perp}$, which completes the proof.



\section{Proof of Theorem \ref{T0906a}}
We convert the complex numbers ${\bf g}_1^H{\bf \hat g}_2$ and ${\bf g}_1^H{\bf \hat g}_2^{\perp}$ into their complex exponential forms such that ${\bf g}_1^H{\bf \hat g}_2=\left|{\bf g}_1^H{\bf \hat g}_2\right|e^{j\angle{\bf g}_1^H{\bf \hat g}_2}$ and ${\bf g}_1^H{\bf \hat g}_2^{\perp}=\left|{\bf g}_1^H{\bf \hat g}_2^{\perp}\right|e^{j\angle{\bf g}_1^H{\bf \hat g}_2^{\perp}}$, and express the optimization variables $\alpha$ and $\beta$ as $\alpha={\kappa}e^{j\angle{\alpha}}$ and $\beta={\iota}e^{j\angle{\beta}}$, respectively, where ${\kappa}, {\iota}\geq 0$. Then, problem \eqref{20180903g} can be equivalently transformed as
\begin{equation}\label{20180905a}
\begin{split}
&\min_{{\kappa}\geq 0, {\iota}\geq 0, \angle{\alpha}, \angle{\beta}}~~{\kappa}^2\left\|{\bf g}_2\right\|^2\\
&~s.t.~\left|{\kappa}\left|{\bf g}_1^H{\bf \hat g}_2\right|e^{j\left(\angle{\alpha}+\angle{\bf g}_1^H{\bf \hat g}_2\right)}+{\iota}\left|{\bf g}_1^H{\bf \hat g}_2^{\perp}\right|e^{j\left(\angle{\beta}+\angle{\bf g}_1^H{\bf \hat g}_2^{\perp}\right)}
\right|^2=\phi\left(R_1\right),\\
&~~~~~~{\kappa}^2+{\iota}^2\leq P_{\rm max}.
\end{split}
\end{equation}

Applying the triangle inequality to the left hand
side (LHS) of the first constraint in problem \eqref{20180905a}, which yields
\begin{equation}\label{20180905b}
\left|{\kappa}\left|{\bf g}_1^H{\bf \hat g}_2\right|e^{j\left(\angle{\alpha}+\angle{\bf g}_1^H{\bf \hat g}_2\right)}+{\iota}\left|{\bf g}_1^H{\bf \hat g}_2^{\perp}\right|e^{j\left(\angle{\beta}+\angle{\bf g}_1^H{\bf \hat g}_2^{\perp}\right)}
\right|^2 \leq \left({\kappa}\left|{\bf g}_1^H{\bf \hat g}_2\right|+{\iota}\left|{\bf g}_1^H{\bf \hat g}_2^{\perp}\right|\right)^2,
\end{equation}
where the equality holds when  $\angle{\alpha}+\angle{\bf g}_1^H{\bf \hat g}_2=\angle{\beta}+\angle{\bf g}_1^H{\bf \hat g}_2^{\perp}=\theta $ with  $\theta$ being any phase.

Notice that the phases $\angle{\alpha}$ and $\angle{\beta}$ only appear in the first constraint of problem \eqref{20180905a}, and the feasible set of ${\kappa}$, ${\iota}$ in problem \eqref{20180905a} can be enlarged if the LHS of the first constraint in problem \eqref{20180905a} is replaced by the right hand side (RHS) of \eqref{20180905b}. Therefore,  the optimal phases $\angle{\alpha}^*$ and $\angle{\beta}^*$ should guarantee that the equality in \eqref{20180905b} can be achieved. Without loss of generality, let $\theta = 0 $, then we have
\begin{equation}\label{20180905c}
\angle{\alpha}^*=-\angle{\bf g}_1^H{\bf \hat g}_2,~{\rm and}~\angle{\beta}^*=-\angle{\bf g}_1^H{\bf \hat g}_2^{\perp}.
\end{equation}

Substituting \eqref{20180905c} into problem \eqref{20180905a}, which can be  simplified as
\begin{equation}\label{20180905d}
\begin{split}
&\min_{{\kappa}\geq 0, {\iota}\geq 0}~~{\kappa}^2\left\|{\bf g}_2\right\|^2\\
&~~~s.t.~~~~\left({\kappa}\left|{\bf g}_1^H{\bf \hat g}_2\right|+{\iota}\left|{\bf g}_1^H{\bf \hat g}_2^{\perp}\right|\right)^2=\phi\left(R_1\right),\\
&~~~~~~~~~~~~{\kappa}^2+{\iota}^2\leq P_{\rm max}.
\end{split}
\end{equation}

We observe that the objective function of problem \eqref{20180905d} is only related
to the optimization variable ${\kappa}$. Then, to solve problem \eqref{20180905d}, we need to find the minimum ${\kappa}$ under the condition that the constraints of problem \eqref{20180905d} are satisfied. Obviously, the minimum value of ${\kappa}$ is zero, i.e., ${\kappa}^*=0$, which can be achieved if the following problem has a feasible solution given by ${\iota}$.
\begin{equation}\label{20180906a}
\begin{split}
& {\rm find}~~~{ {\iota}}\\
&~s.t.~~{\iota}^2\left|{\bf g}_1^H{\bf \hat g}_2^{\perp}\right|^2=\phi\left(R_1\right),~{\iota}^2\leq P_{\rm max}.
\end{split}
\end{equation}

It is easy to verify that problem \eqref{20180906a} is feasible if and only if $\phi\left(R_1\right) \leq P_{\rm max}\left|{\bf g}_1^H{\bf \hat g}_2^{\perp}\right|^2$, and the corresponding optimal solution to problem \eqref{20180906a} is ${\iota}^*=\sqrt{\frac{\phi\left(R_1\right)}{\left|{\bf g}_1^H{\bf \hat g}_2^{\perp}\right|^2}}$.
Therefore, we can conclude that, if the condition $\phi\left(R_1\right) \leq P_{\rm max}\left|{\bf g}_1^H{\bf \hat g}_2^{\perp}\right|^2$ is satisfied, the optimal solutions to problem \eqref{20180903g} are
\begin{equation}\label{20180906b}
\alpha^*={\kappa}^*e^{j\angle{\alpha}^*}=0,
\end{equation}
\begin{equation}\label{20180906bsdsd} ~\beta^*={\iota}^*e^{j\angle{\beta}^*}=\sqrt{\frac{\phi\left(R_1\right)}{\left|{\bf g}_1^H{\bf \hat g}_2^{\perp}\right|^2}}e^{-j\angle{\bf g}_1^H{\bf \hat g}_2^{\perp}}.
\end{equation}

In what follows, we focus on solving problem \eqref{20180905d} for the case that $\phi\left(R_1\right) > P_{\rm max}\left|{\bf g}_1^H{\bf \hat g}_2^{\perp}\right|^2$.

By contradiction, it is easy to verify that if $\phi\left(R_1\right) > P_{\rm max}\left|{\bf g}_1^H{\bf \hat g}_2^{\perp}\right|^2$, then the optimal solution to problem \eqref{20180905d} should satisfy ${\kappa}^{*^2}+{\iota}^{*^2}=P_{\rm max}$.
%
Thus, in this case,  problem \eqref{20180905d} is equivalent to the following problem.
\begin{equation}\label{20180906b}
\begin{split}
&\min_{{\kappa}\geq 0, {\iota}\geq 0}~~{\kappa}^2\left\|{\bf g}_2\right\|^2\\
&~~~s.t.~~~~~\left({\kappa}\left|{\bf g}_1^H{\bf \hat g}_2\right|+{\iota}\left|{\bf g}_1^H{\bf \hat g}_2^{\perp}\right|\right)^2=\phi\left(R_1\right),\\
&~~~~~~~~~~~~{\kappa}^2+{\iota}^2= P_{\rm max}.
\end{split}
\end{equation}

Using the relation ${\iota}=\sqrt{P_{\rm max}-{\kappa}^2}$, we can combine the two constraints in problem \eqref{20180906b} into one single constraint $\left({\kappa}\left|{\bf g}_1^H{\bf \hat g}_2\right|+\sqrt{P_{\rm max}-{\kappa}^2}\left|{\bf g}_1^H{\bf \hat g}_2^{\perp}\right|\right)^2=\phi\left(R_1\right)$, and thus problem \eqref{20180906b} is simplified as the following single-variable optimization problem.
\begin{equation}\label{20180906c}
\begin{split}
&\min_{{\kappa}\geq 0}~~{\kappa}^2\left\|{\bf g}_2\right\|^2\\
&~~s.t.~~{\kappa}\left|{\bf g}_1^H{\bf \hat g}_2\right|+\sqrt{P_{\rm max}-{\kappa}^2}\left|{\bf g}_1^H{\bf \hat g}_2^{\perp}\right|=\sqrt{\phi\left(R_1\right)}.
\end{split}
\end{equation}

It is interesting to observe that the unique equality constraint in problem \eqref{20180906c} can be equivalently transformed to the following quadratic equation with respect to ${\kappa}$:
\begin{equation}\label{20180906d}
a_0{\kappa}^2-2\left|{\bf g}_1^H{\bf \hat g}_2\right|\sqrt{\phi\left(R_1\right)}{\kappa}+\phi\left(R_1\right)-\left|{\bf g}_1^H{\bf \hat g}_2^{\perp}\right|^2P_{\rm max}=0,
\end{equation}
where $a_0=\left|{\bf g}_1^H{\bf \hat g}_2\right|^2+\left|{\bf g}_1^H{\bf \hat g}_2^{\perp}\right|^2$.

 It is easy to check that the discriminant of the quadratic equation defined
in \eqref{20180906d} is greater than zero, i.e.,  $a_0P_{\rm max}-\phi\left(R_1\right)>0$. Thus, it has two distinct real roots and are respectively given by
\begin{equation}\label{20180906e}
{\kappa}^{1*}=\frac{\left|{\bf g}_1^H{\bf \hat g}_2\right|\sqrt{\phi\left(R_1\right)}-\left|{\bf g}_1^H{\bf \hat g}_2^{\perp}\right|\sqrt{a_0P_{\rm max}-\phi\left(R_1\right)}}{a_0},
\end{equation}
\begin{equation}\label{20180906f}
{\kappa}^{2*}=\frac{\left|{\bf g}_1^H{\bf \hat g}_2\right|\sqrt{\phi\left(R_1\right)}+\left|{\bf g}_1^H{\bf \hat g}_2^{\perp}\right|\sqrt{a_0P_{\rm max}-\phi\left(R_1\right)}}{a_0}.
\end{equation}

Next, we show that ${\kappa}^{1*}\geq 0$ by contradiction. Suppose that ${\kappa}^{1*} <0$. Then,  we have
\begin{equation}\label{20180906g}
\frac{\left|{\bf g}_1^H{\bf \hat g}_2\right|\sqrt{\phi\left(R_1\right)}-\left|{\bf g}_1^H{\bf \hat g}_2^{\perp}\right|\sqrt{a_0P_{\rm max}-\phi\left(R_1\right)}}{a_0} < 0.
\end{equation}

Note that the inequality in \eqref{20180906g} can be directly simplified as $\phi\left(R_1\right) < P_{\rm max}\left|{\bf g}_1^H{\bf \hat g}_2^{\perp}\right|^2$, which  contradicts the above assumption that $\phi\left(R_1\right) \geq P_{\rm max}\left|{\bf g}_1^H{\bf \hat g}_2^{\perp}\right|^2$. Thus, ${\kappa}^{1*}\geq 0$ is true.

As a result, ${\kappa}^{1*}$ and ${\kappa}^{2*}$ are two feasible solutions to problem \eqref{20180906c}. Obviously, ${\kappa}^{1*}$ is the optimal solution to problem \eqref{20180906c} because it leads to a smaller objective value than ${\kappa}^{2*}$. Thus, the optimal solutions to problem \eqref{20180906b} are

\begin{equation}\label{20180906h}
{\kappa}^*={\kappa}^{1*},~~{\iota}^*=\sqrt{P_{\rm max}-{\kappa}^{*^2}}.
\end{equation}

Combining \eqref{20180905c} and \eqref{20180906h}, we obtain that, if $\phi\left(R_1\right) \geq P_{\rm max}\left|{\bf g}_1^H{\bf \hat g}_2^{\perp}\right|^2$, the solutions to problem  \eqref{20180903g} are
\begin{equation}\label{20180906i}
\alpha^*={\kappa}^{1*}e^{-j\angle{\bf g}_1^H{\bf \hat g}_2},~~\beta^*={\iota}^*e^{-j\angle{\bf g}_1^H{\bf \hat g}_2^{\perp}}.
\end{equation}

Based on \eqref{20180906b} and \eqref{20180906i}, we obtain the Theorem \ref{T0906a}, and the proof is completed.

\section{Proof of Proposition \ref{P20181002a}}
For convenience, we define $a_1=P_1\left|h_{1,1}\right|^2$, $a_2=P_2\left|h_{2,2}\right|^2$, $c_1={\tilde \sigma}_1^2$, $c_2={\tilde \sigma}_2^2$, $b_1=\left|{\bf g}_1^H{\bf \hat g}_2\right|$,  $b_2=\left|{\bf g}_1^H{\bf \hat g}_2^{\perp}\right|$, $b_3=b_1^2+b_2^2$, and $b_4=\frac{\left\|{\bf g}_2\right\|^2}{b_3^2}$. Then, $f_{\rm max}\left(R_1\right)$ can be re-expressed as
\begin{equation}\label{20181002a}
f_{\rm max}\left(R_1\right)\!=\!{\rm log}\left(1+\Gamma\left(R_1\right)\right),
\end{equation}
where $\Gamma\left(R_1\right)=\frac{a_2b_3^2}{b_4\left(b_1\sqrt{\phi\left(R_1\right)}\!-\!b_2\sqrt{b_3P_{\rm max}\!-\!\phi\left(R_1\right)}\right)^2\!+\!c_2}$, with $\phi\left(R_1\right)=\frac{a_1}{2^ {R_1}-1}-c_1$.

By taking into account the monotonicity of the logarithm function, we know that the monotonicity of $f_{\rm max}\left(R_1\right)$ is similar to that of $\Gamma\left(R_1\right)$. Therefore, we turn to check the monotonicity of $\Gamma\left(R_1\right)$. Specifically, the first-order derivative of $\Gamma\left(R_1\right)$ is given by
\begin{equation}\label{20181002b}
\Gamma^{\prime}\left(R_1\right)=\frac{2^{R_1}{\rm ln}\left(2\right)a_1a_2b_3^2b_4\left(\frac{b_1}{\sqrt{\frac{a_1}{ 2^ {R_1}-1}-c_1}}+\frac{b_2}{\sqrt{c_1+b_3P_{\rm max}-\frac{a_1}{ 2^ {R_1}-1}}}\right)\varphi\left(R_1\right)}{\left(2^ {R_1}-1\right)^2\left(b_4\varphi\left(R_1\right)^2+c_2\right)^2},
\end{equation}
where $\varphi\left(R_1\right)=b_1\sqrt{\frac{a_1}{ 2^ {R_1}-1}-c_1}-b_2\sqrt{c_1+b_3P_{\rm max}-\frac{a_1}{ 2^ {R_1}-1}}$.

Obviously, whether $\Gamma^{\prime}\left(R_1\right)$ is positive
or not depends only on $\varphi\left(R_1\right)$. By contradiction, it is easy to check that $\varphi\left(R_1\right)$ is positive when $R_1 <R_1^{\rm ZF2}$.  Hence, we know that $\Gamma^{\prime}\left(R_1\right)>0$  when $R_1 <R_1^{\rm ZF2}$. As a result, we arrive at Proposition \ref{P20181002a},  which completes the proof.

\section{Proof of Theorem \ref{T20180907ass}}
To prove Theorem \ref{T20180907ass}, we first apply the triangle inequality to the objective function of problem \eqref{E17ssssss}, which yields
\begin{equation}\label{20181001a}
\left|\mu{\bf g}_2^H{\bf \hat g}_1+\nu{\bf g}_2^H{\bf \hat g}_1^{\perp}\right|^2
\leq \left(\left|\mu\right|\left|{\bf g}_2^H{\bf \hat g}_1\right|+\left|\nu\right|\left|{\bf g}_2^H{\bf \hat g}_1^{\perp}\right|\right)^2,
\end{equation}
where the equality holds when $\angle \mu +\angle{\bf g}_2^H{\bf \hat g}_1=\angle \nu +\angle{\bf g}_2^H{\bf \hat g}_1^{\perp}$. Notice that the phases of $\mu$ and $\nu$ do not affect the feasible set of problem \eqref{E17ssssss}. Therefore, the optimal solution to problem \eqref{E17ssssss} should make the equality in \eqref{20181001a} hold. Without loss of generality, we let
\begin{equation}\label{20181001b}
\angle \mu^*=-\angle{\bf g}_2^H{\bf \hat g}_1, ~{\rm and}~ \angle \nu^* =-\angle{\bf g}_2^H{\bf \hat g}_1^{\perp}.
\end{equation}

By contradiction, we can easily verify that the power constraint in problem \eqref{E17ssssss} should be active at the optimal solution, i.e.,
$\left|\mu^*\right|^2+\left|\nu^*\right|^2 = P_{\rm max}$. Then, we have
\begin{equation}\label{E17ssesssss}
\left|\mu^*\right|^2=\frac{\phi\left(R_1\right)}{\left\|{\bf g}_1\right\|^2}~{\rm and}~\left|\nu^*\right|^2=P_{\rm max}-\frac{\phi\left(R_1\right)}{\left\|{\bf g}_1\right\|^2}.
\end{equation}

From \eqref{20181001b} and \eqref{E17ssesssss}, it follows that
\begin{equation}\label{E17ssesssss1}
\mu^*=\sqrt{\frac{\phi\left(R_1\right)}{\left\|{\bf g}_1\right\|^2}}e^{-j\angle{\bf g}_2^H{\bf \hat g}_1}, \nu^*=\sqrt{P_{\rm max}-\frac{\phi\left(R_1\right)}{\left\|{\bf g}_1\right\|^2}}e^{-j\angle{\bf g}_2^H{\bf \hat g}_1^{\perp}}
 \end{equation}

Then, according to  Lemma \ref{P20180725} and \eqref{E17ssesssss1}, we can obtain the optimal solution to problem \eqref{20180907a}, i.e., ${\bf w}_{\rm opt}=\mu^*{\bf \hat g}_1+\nu^* {\bf \hat g}_1^{\perp}$. By substituting ${\bf Q}={\bf w}_{\rm opt}{\bf w}_{\rm opt}^H$ to the objective function of problem \eqref{20180903b}, the closed-form expression of $f_{\rm min}\left(R_1\right)$ is thus given by
\begin{equation}\label{20181001cs}
f_{\rm min}\left(R_1\right) ={\rm log}\left(1+\frac{P_2\left|h_{2,2}\right|^2}{\left(\left|\mu^*\right|\left|{\bf g}_2^H{\bf \hat g}_1\right|+\left|\nu^*\right|\left|{\bf g}_2^H{\bf \hat g}_1^{\perp}\right|\right)^2+{\tilde \sigma}_2^2}\right).
\end{equation}

Hence, the proof is completed.

\end{document}